\documentclass[namedreferences]{SolarPhysics}
\usepackage[optionalrh]{spr-sola-addons} 
\usepackage{graphicx}                    
\usepackage{color}                       
\usepackage{url}                         

\begin{document}

\begin{article}

\begin{opening}

\title{Modeling the Longitudinal Asymmetry in Sunspot Emergence --- the Role of the Wilson Depression}

%
\author{F.~\surname{Watson}$^{1}$\sep
        L.~\surname{Fletcher}$^{1}$\sep
       S.~\surname{Dalla}$^{2}$ \sep
       S.~\surname{Marshall}$^{3}$ \sep    
       }

%

%
  \institute{$^{1}$ Department of Physics and Astronomy, University of Glasgow, Glasgow G12 8QQ, U.K.
                     email: \url{f.watson@astro.gla.ac.uk} email: \url{lyndsay@astro.gla.ac.uk}\\ 
                $^{2}$ Jeremiah Horrocks Institute for Astrophysics and Supercomputing, University of Central Lancashire, Preston PR1 2HE, UK.  email: \url{sdalla@uclan.ac.uk} \\
                $^{3}$ Department of Electronic and Electrical Engineering, University of Strathclyde, Glasgow G1 1XW.  email: \url{s.marshall@eee.strath.ac.uk}  }

\begin{abstract}
The distributions of sunspot longitude at first appearance and at disappearance display an east-west asymmetry, that results from a reduction in visibility as one moves from disk centre to the limb. To first order, this is explicable in terms of simple geometrical foreshortening. However, the centre to limb visibility variation is much larger than that predicted by foreshortening. Sunspot visibility is known also to be affected by the Wilson effect - the apparent `dish' shape of the sunspot photosphere caused by the temperature-dependent variation of the geometrical position of the $\tau = 1$ layer. In this paper we investigate the role of the Wilson effect on the sunspot appearance distributions, deducing a mean depth for the umbral $\tau = 1$ layer of 500-1500 km. This is based on the comparison of observations of sunspot longitude distribution and Monte-Carlo simulations of sunspot appearance using different models for spot growth rate, growth time and depth of Wilson depression.
\end{abstract}

%

\end{opening}

%
 \section{Introduction and Previous Work}\label{section:introduction} 
The appearance of sunspots on the solar photosphere has long been an indicator of solar activity. They are observed as dark, cool patches on the surface of the Sun. The high magnetic fields in a sunspot cause a disruption in the convection layer directly below the spot and prevent the transfer of heat into the spot region causing the dark spots that are seen. 

Using the USAF/Mt. Wilson sunspot catalogue data between 1 December 1981 and 31 December 2005, processed using AstroGrid workflows, \inlinecite{2008A&A...479L...1D} studied the appearance and disappearance of sunspots from the solar disk. The longitude at which a sunspot first appears depends on solar rotation which carries the spot onto and off the observable hemisphere,  the manner of growth and decay of the sunspots, and the minimum area required for a positive detection, which is specific to the  instrument and observer. All factors affecting the latter are described by the `visibility function',  a curve that gives the minimum required sunspot area for the spot to be visible at a given longitude. A sunspot is observable when its area at a given longitude exceeds the value of the visibility function at that longitude. The combination of the evolution (rotation, growth, decay) of a sunspot and the visibility function determines the distribution of longitudes at which a spot is observed for the first or last time. Somewhat counterintuitively, a symmetric visibility function with minimum at disk centre (the location where our view of sunspots is best), results in an east-west asymmetry in observed location of new spot emergence. \inlinecite{1907MNRAS..67..451M} was the first to observe such an asymmetry in sunspot data, while \inlinecite{schuster1911} and \inlinecite{1939AN....269...48M} provided the interpretation of this phenomenon as a visibility effect. Applying the methodology of \inlinecite{schuster1911} to the USAF/Mount Wilson data, \inlinecite{2008A&A...479L...1D}  determined the visibility function for different values of the spot growth rate, and showed that for the  range of feasible growth rates the visibility curve is inconsistent with that predicted solely from geometrical foreshortening. The strong centre to limb variation in visibility observed by \inlinecite{2008A&A...479L...1D} is as yet not fully explained.

It is therefore natural to investigate other factors which affect the sunspot visibility, and probe the possible diagnostic potential of the observed sunspot distributions. The first two factors that spring to mind, in the sense that these are sunspot properties that have been known for centuries, are the internal umbra/penumbra structure, and the Wilson effect (sometimes known as the Wilson depression). 

Based on their `dished' appearance at the solar limb, in 1769 A. Wilson theorised that sunspot visible radiation emerged from a layer deeper than the surrounding quiet photosphere, with the umbral radiation emerging from deeper than the penumbral radiation. This is now understood at least qualitatively in terms of the temperature structure of the sunspot photosphere and some detailed modelling work and observational interpretation of single well-observed sunspots also exist \cite{2004A&A...422..693M}. Sunspots are cooler than their surroundings because of the strong inhibition of convection due to their magnetic field, resulting in the geometrical depth corresponding to an optical depth of $\tau = 1$ being below the level of the surrounding photosphere. The importance for our problem is that for a sunspot on the solar limb, the $\tau=1$ geometric layer of the surrounding photosphere nearest the observer can, for a range of viewing angles, substantially occlude the $\tau=1$ geometrical layer of the penumbra/umbra. The result is that the apparent width of the penumbra/umbra on the side of the spot closer to the disk centre is smaller than that of the penumbra on the limbward side of the spot. 

 Geometrical projection also affects the observed spot area. The width of the various parts of the spot decreases, with the limbward side of the spot showing a greater decrease in width as it is closer to the limb. However, the occluding effect of the diskward photosphere also renders smaller the detectable penumbral area. In this paper we adopt a very simple model for the sunspot, with a rectangular profile of $\tau = 1$ (see Figure~\ref{fig:wdcartoon}), which is, nonetheless compatible with the basis on which we construct our sunspot sample. The free parameters of this model are the depth of the $\tau = 1$ layer, the spot growth rate and the area distribution of spots at emergence. 
The depth of the $\tau = 1$ layer (also known as the Wilson depression) is difficult to measure due to the constant evolution and changes in the width of the sunspot regions. Values range from 400-800 km (\opencite{1972SoPh...26...52G}) right up to 1500-2100 km (\opencite{1974SoPh...35..105P}), though most common estimates are around 1000 km. One goal of this paper is to determine values of the Wilson depression which are, within the limits of the model assumptions, consistent with the sunspot asymmetry data.

The structure of the paper is as follows. In Section~\ref{sect:distributions} we discuss the sunspot dataset used, how it is obtained and the generation of the sunspot appearance distributions. The forward models using Monte-Carlo simulations, and their comparison with the observational data are presented in Section~\ref{sect:monte} and Section~\ref{sect:testing} respectively, and the results of inclusion of the Wilson effect in Section~\ref{sect:wilson}. We then search for an optimum value of the depth of the $\tau = 1$ layer in sunspot umbrae in Section~\ref{sect:optimum} and finish with Discussion and Conclusions.

\section{Longitude Distributions of Sunspot Emergence and Disappearance}\label{sect:distributions}
For this work, we make use of image processing techniques to automatically detect sunspots and derive new distributions of sunspot appearances. The reason for doing this is that we wish to record the emergence of individual spots as opposed to the sunspot groups which are recorded in the Mt. Wilson catalogue. In this paper we use a simplified model for the Wilson effect in circular spots, and it is more meaningful to compare this with single spots than with groups.

Our sunspot identification algorithm uses tools from mathematical morphology. Mathematical morphology is a nonlinear image processing technique developed by \inlinecite{matheron1975} and \inlinecite{serra1982} that uses shape and structure in a digital image to analyse various aspects of the image. Matheron and Serra developed the theory with binary images but it has since been extended to grayscale and colour. All of the operations of mathematical morphology can be broken down into two operators, erosion and dilation, both of which will be explained in this paper. In an erosion or dilation the image is probed with a shape known as the structuring element as will be described later.  Mathematical morphology techniques have also recently been used by \inlinecite{2006A&A...457..729M} for detecting and removing cosmic ray noise from coronagraph images, and by \inlinecite{2008SoPh..tmp..124C} for sunspot detection. 

The approach we take is as follows: we use SOHO MDI level 1.5 continuum data \cite{1995SoPh..162..129S}. The FITS files are read into MATLAB and a morphological top-hat transform (\opencite{DoughertyHandsOn}) is applied to the data. For this to work, a structuring element must be chosen that is larger than the features to be detected. The top-hat transform is defined as
\begin{equation}
 F = O - (O \circ SE)
\end{equation}
where $F$ is the final transformed image, $O$ is the original image and $\circ$ corresponds to a morphological opening operation. An opening operation is defined as an erosion followed by a dilation. The $SE$ is the structuring element used as the probe in the transform (see \inlinecite{DoughertyImageProc} for more details). Note that the minus sign in equation 1 corresponds to a pixel by pixel subtraction.

\begin{figure}
\centering
\includegraphics[width=0.7\textwidth]{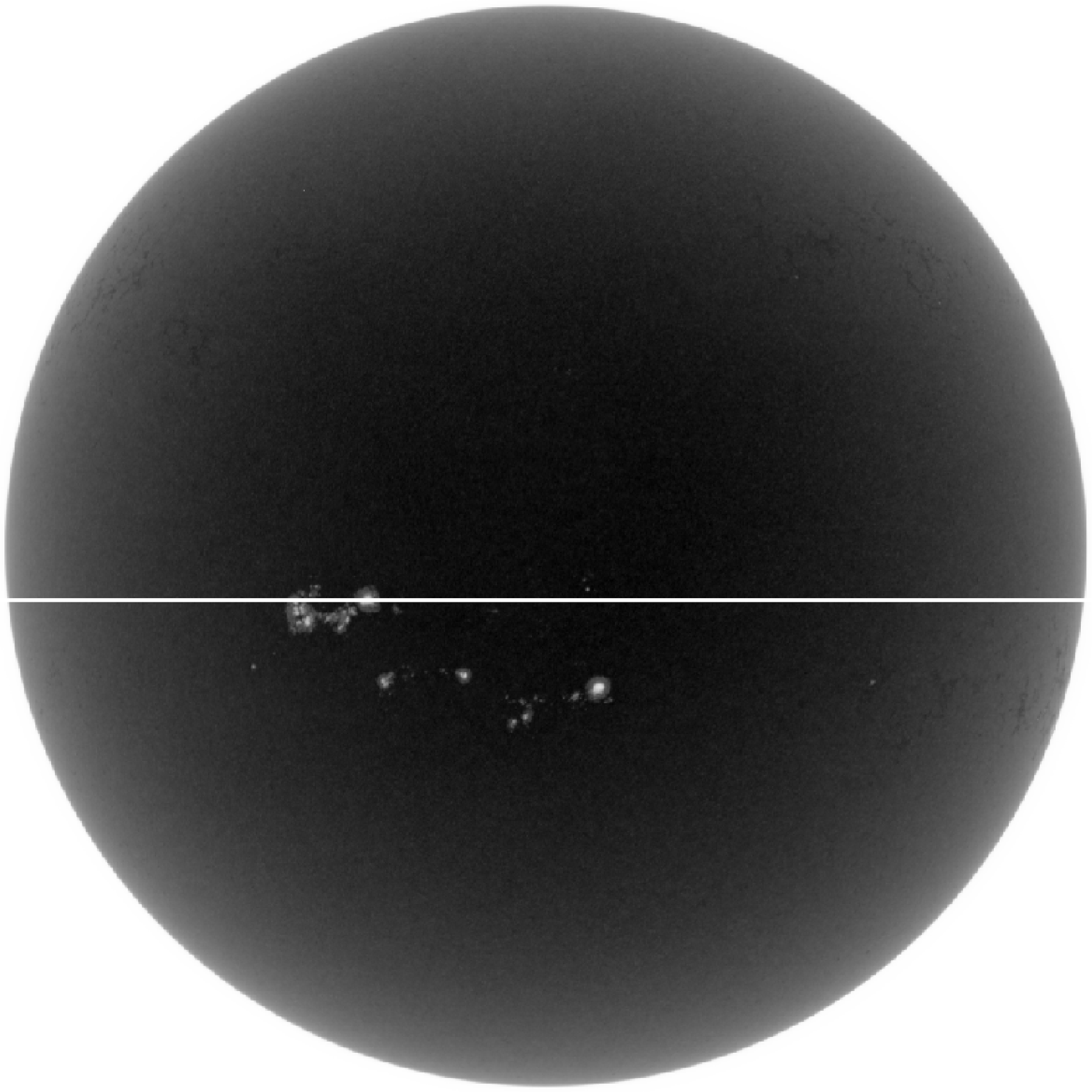}
\centering
\includegraphics[width=0.95\textwidth]{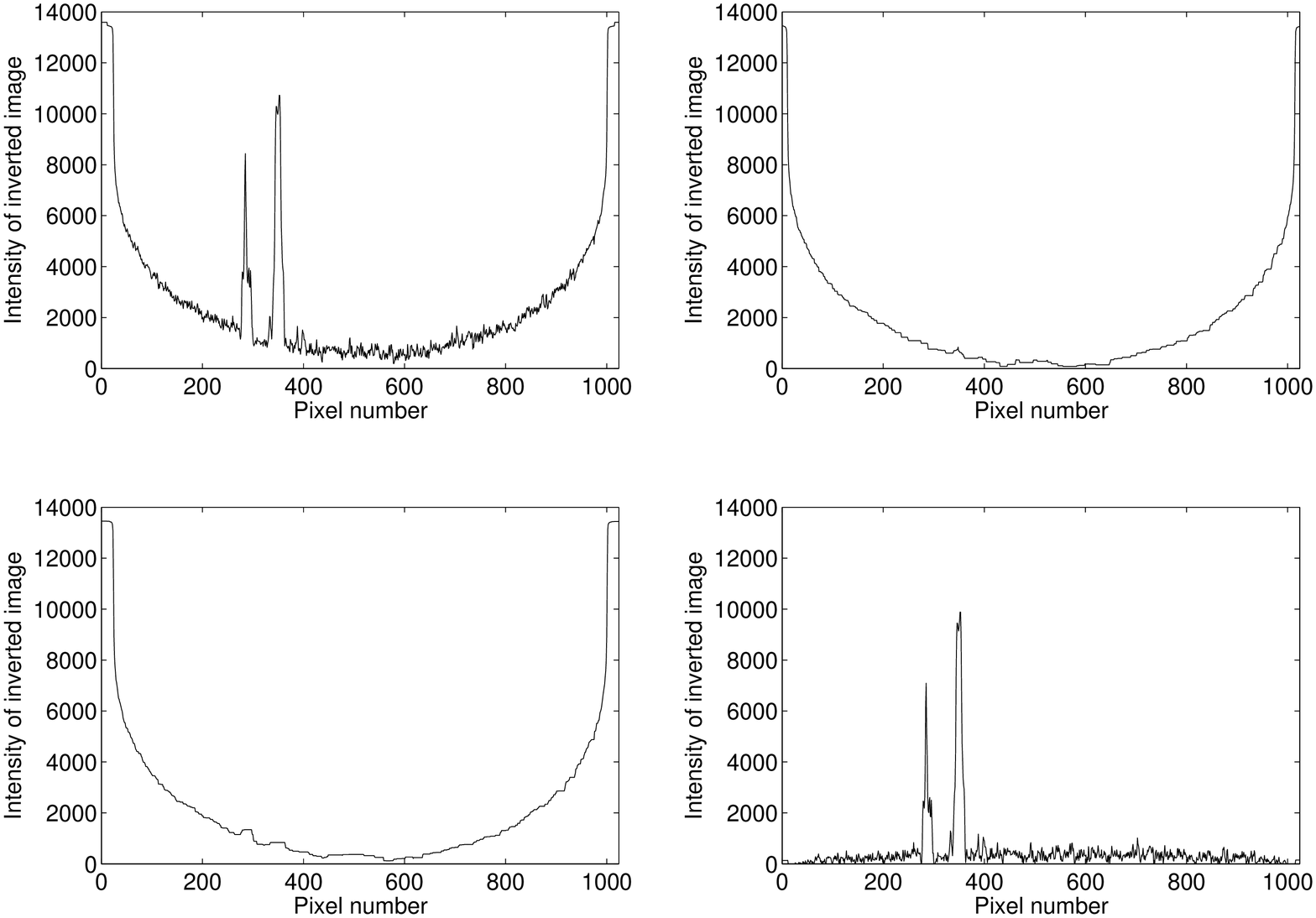}
\caption{The various stages of the top-hat transform detection algorithm. (top) - original image after inversion. (top left panel) - intensity values of the row marked by the line on the inverted original were plotted. (top right panel) - a morphological erosion was performed. Note how the plot is much smoother as the structuring element cannot fit into the variations in intensity of the solar surface. (bottom left panel) - a morphological dilation was performed with the same structuring element as in the erosion. This returns approximately the same intensity profile as the original but now the sunspot peaks have been removed. (bottom right panel) the dilated profile is subtracted from the original to leave the sunspot peaks intact.}\label{fig:tophat}
\end{figure}

 To explain how the transform works, it is more instructive to take a single row of pixels from an image and plot the intensity as a function of pixel number in that row to see the effects of each stage of the transform (Figure~\ref{fig:tophat}).

The images are initially inverted for illustration so that the sunspots appear as bright areas and as peaks on the graph (Figure~\ref{fig:tophat} top and top-left panel). This is done because the top-hat transform detects the intensity peaks in an image. The structuring element used to probe this profile is a circle having a diameter of 28 MDI pixels. The inverted intensity profile is morphologically eroded (Figure~\ref{fig:tophat} top-right panel), which is a process used to remove peaks from a profile. This can be thought of as moving the structuring element along underneath the profile (ensuring it is always touching the profile) and marking out the path of the centre of the structuring element. As the structuring element is chosen to be larger than the width of the sunspot peaks, it cannot `fit into' those peaks and so their size is greatly reduced. This profile is then morphologically dilated (Figure~\ref{fig:tophat} bottom-left panel). This process is the dual transform to erosion and can be thought of as moving the structuring element \textit{above} the profile and tracing the centre point (this is necessary as the profile is now 14 units of intensity below the original). Doing this gives a very similar profile to that of the original image but \emph{without} the sunspot peaks present. 

This profile is then subtracted from the original (Figure~\ref{fig:tophat} bottom-right panel) to leave sunspot candidates intact. Then an intensity threshold is applied and any peaks above this threshold and larger than a critical area are recorded as sunspots. This method is very effective as it ignores any slow variations in intensity such as the limb darkening effect observed in the data due to the structuring element being smaller than the scale of the variation. In addition to this, the algorithm is efficient and could process an image in 3-4 seconds. This method did pick up some stray pixels due to dead areas on the CCD or cosmic ray hits and so a median filter was applied to smooth out those pixels. The algorithm is applied to all rows of the image to build up a 3-dimensional intensity map that shows the sunspot peaks. To remove all candidates that are not sunspots, areas less than 30 MDI pixels are removed. This corresponds to an area of 20 millionths of a solar hemisphere.

The method was tested against `ground truth' images, these being a set of eight MDI images from the interval January 1998 to October 2002, in which a human observer had marked all of the pixels judged to belong to sunspots. This does introduce an element of subjectivity into the process, but even the best feature recognition algorithms are only as good as a human observer. This leads to the same question of reliability and `personal equation' that dogs spot catalogues compiled by hand, but we believe is an improvement on the Mt. Wilson data for the purposes of our investigation, as it rapidly identifies individual spots instead of groups. This is discussed further in Section~\ref{sect:optimum}. The top hat transform was found to recover 77\% of the sunspot pixels, (and 100\%  of the individual spots), and generated roughly 4\% false positive pixels. False pixels consisted of enhancements to the areas of true spots.

\begin{figure*}
   \centerline{\hspace*{0.015\textwidth}
               \includegraphics[width=0.5\textwidth,clip=]{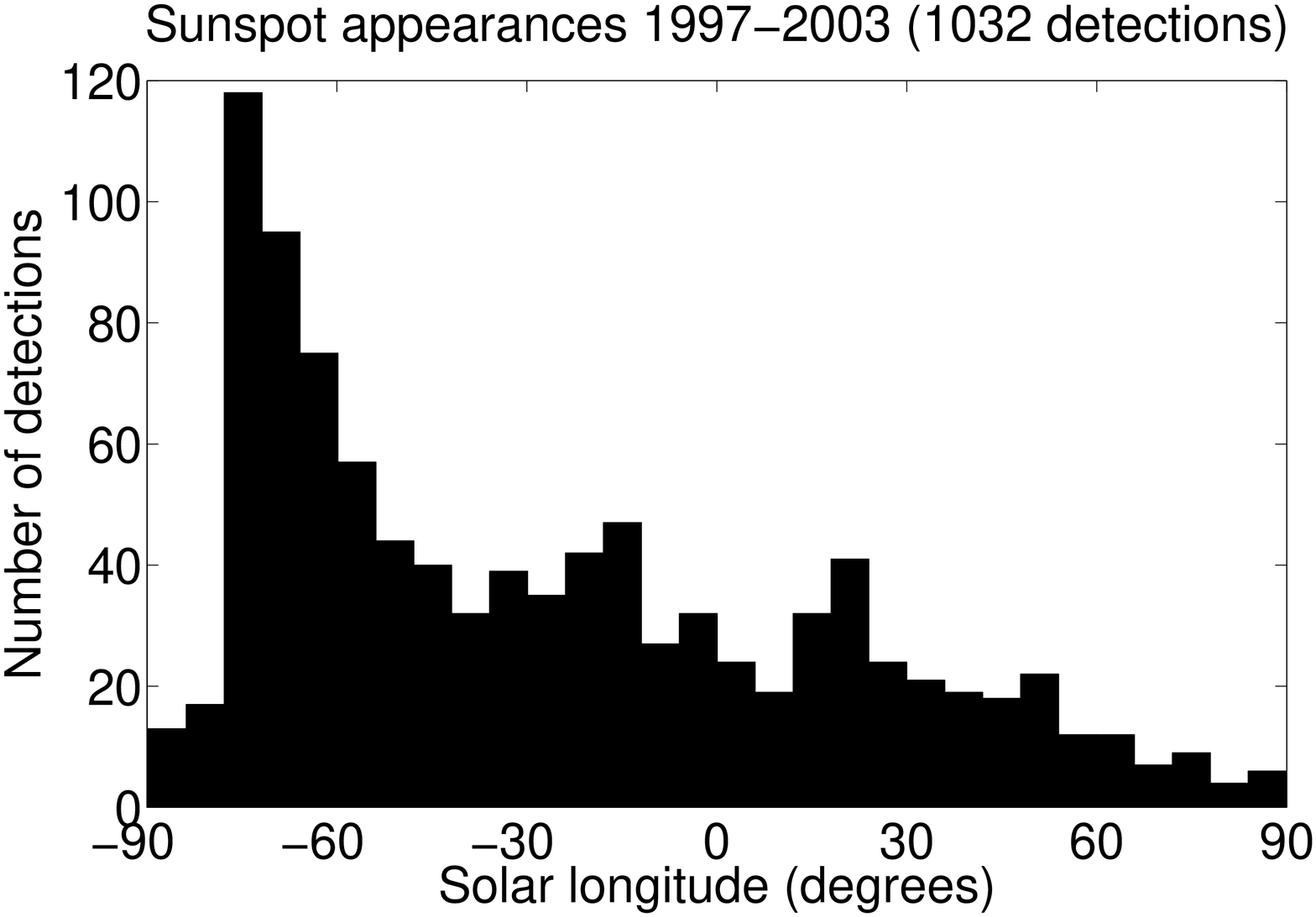}
               \hspace*{0\textwidth}
               \includegraphics[width=0.5\textwidth,clip=]{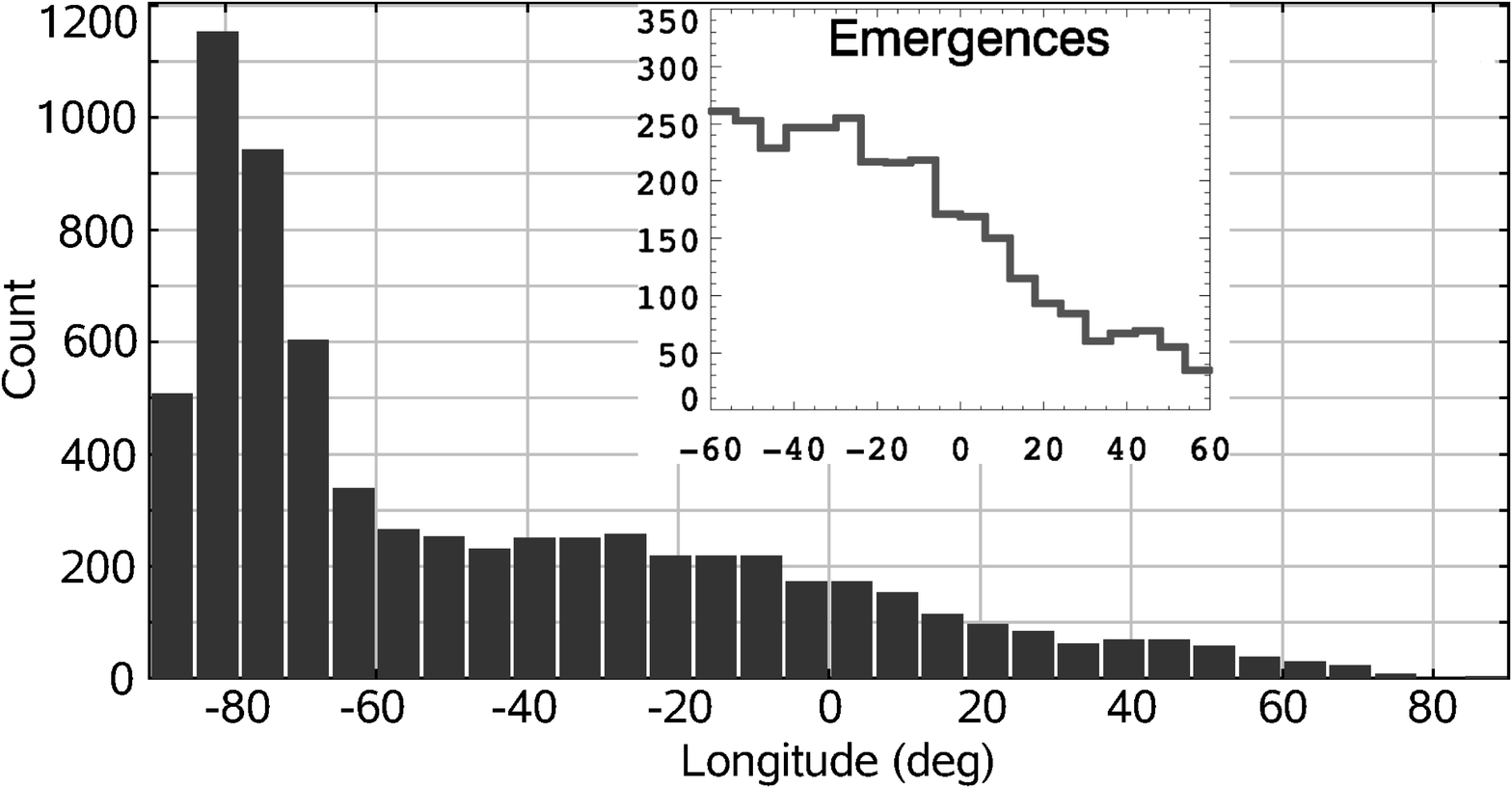}
              }
\caption{The histogram of sunspot appearance longitudes detected in the present work (left), and based on the USAF/Mount Wilson catalogue (right).}\label{fig:histograms}
\end{figure*}

The transform is then run on a series of MDI white light images between January 1997 and March 2003, spanning the last solar maximum, and taken at intervals 6 hours apart. Spots are identified in each image, and the MDI pixel co-ordinates of the centroid of each spot determined using the values of the boundary pixels. The pixel co-ordinates are then converted to solar heliographic co-ordinates using the known transformations in SolarSoft \cite{1998SoPh..182..497F}, correcting for SOHO position. These heliographic co-ordinates can be rotated in longitude by 6 hours (the time interval between successive MDI continuum images) using the empirical rotation model of \inlinecite{1990SoPh..130..295H} and corrected for synodic viewing. The new co-ordinate positions are compared with the heliographic co-ordinate positions of all spots present in the next image. The spots can be assumed to be the same if the centroid positions are within two degrees of one another. This is large enough to allow for any drift of the centroid of the spots over six hours and small enough to prevent confusion with other spots in the same group. Spots which either appear or disappear between images can easily be identified in this manner.  The distribution of the longitude of sunspot appearances is presented in Figure~\ref{fig:histograms}, together with the distribution reported by \inlinecite{2008A&A...479L...1D}.

There are differences in the two results although the same overall trends are present. When subjected to a Kolmogorov-Smirnov (KS) test (see Section~\ref{sect:monte}), the two distributions were shown to be consistent at a 5\% significance level. This means that there is a 5\% chance that an agreement as good as, or better than, the agreement found can be generated by random chance. As a result, a lower significance level is better in this test. The larger fluctuations in the left histogram of Figure~\ref{fig:histograms} are a result of a lower sample size.

\section{Monte-Carlo Simulations of Sunspot Emergence and Growth}~\label{sect:monte}
In this section, we investigate whether simple models for spot evolution can explain the observed distributions. To model the sunspot emergence asymmetry, 30000 sunspots are generated uniformly around the solar equator with a power law in initial spot area to generate a set of 30000 random spot areas (note, the same procedure using a single value of initial spot area does not give satisfactory results in terms of the final distribution). Solar rotation is then included in the model with a synodic solar rotation rate of 13.199$^\circ$ per day. This differs from the sidereal rotation rate of the sun at its equator (14.18$^\circ$ per day) due to the images being taken by SOHO which orbits in the direction of solar rotation during the imaging, so that the apparent rotation speed is lower \cite[has a full treatment for more detail]{1995SoPh..159..393R}.

Also taken into account is the effect of geometric foreshortening. As a spot moves closer to the limb of the Sun, the apparent area of that spot from the point of view of the observing spacecraft is reduced. This effect can be modelled simply by saying that
\begin{equation}
A_{\mathrm{app}} = A_0 \cos(\delta)\cos(\lambda)
\end{equation}
where $A_{\mathrm{app}}$ is the apparent area of a spot of area $A_0$ observed to be at a solar latitude of $\delta$ and a longitude of $\lambda$.

The effect of the growth of spots is then included in the models, taking the form of growth rate which is proportional to the area of a given spot at a given time (\opencite{1992SoPh..137...51H}). We also assume that spot area at first appearance is a power law $N(A) = N_0 A^{-p}$ and vary power law index $p$ (see Section~\ref{sect:optimum} for results obtained using a lognormal distribution). Multiple simulations are run with different parameter combinations and spot appearance positions binned into histograms for comparison with observed data.

To determine which set of parameter values gives the best fit to observations, we use the KS test. It is particularly suitable as it is independent of the form of the spot appearance distributions and gives the likelihood that the distributions were both drawn from the same population. The relevant output from the KS test is a number commonly referred to as the $D$-value and is given by

\begin{equation}
 D = \mathrm{sup}|F(x) - S(x)|
\end{equation}

where $F(x)$ and $S(x)$ are the cumulative distribution functions of the modeled and observed sunspot emergence distributions. A lookup table can then be used to determine whether the two distributions are likely to be from the same population. We use a 5\% level of significance in our tests as described in Section~\ref{sect:distributions}. Crucially, a \emph{lower} $D$-value is better as it means the differences between the model and observations are smaller.

\begin{figure}
\centering
 \begin{tabular}{c c}

  \includegraphics[width=0.45\textwidth]{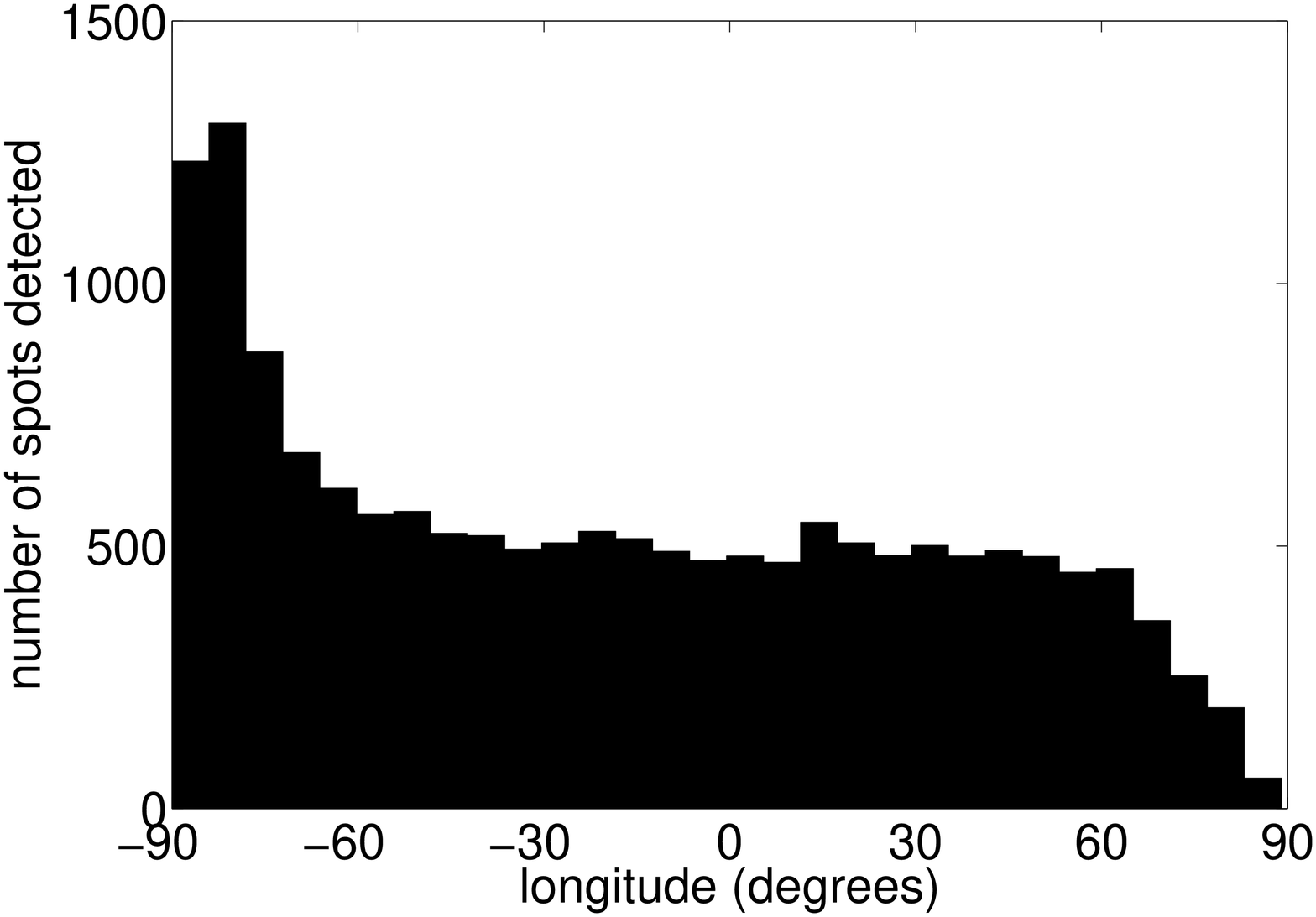} & \includegraphics[width=0.45\textwidth]{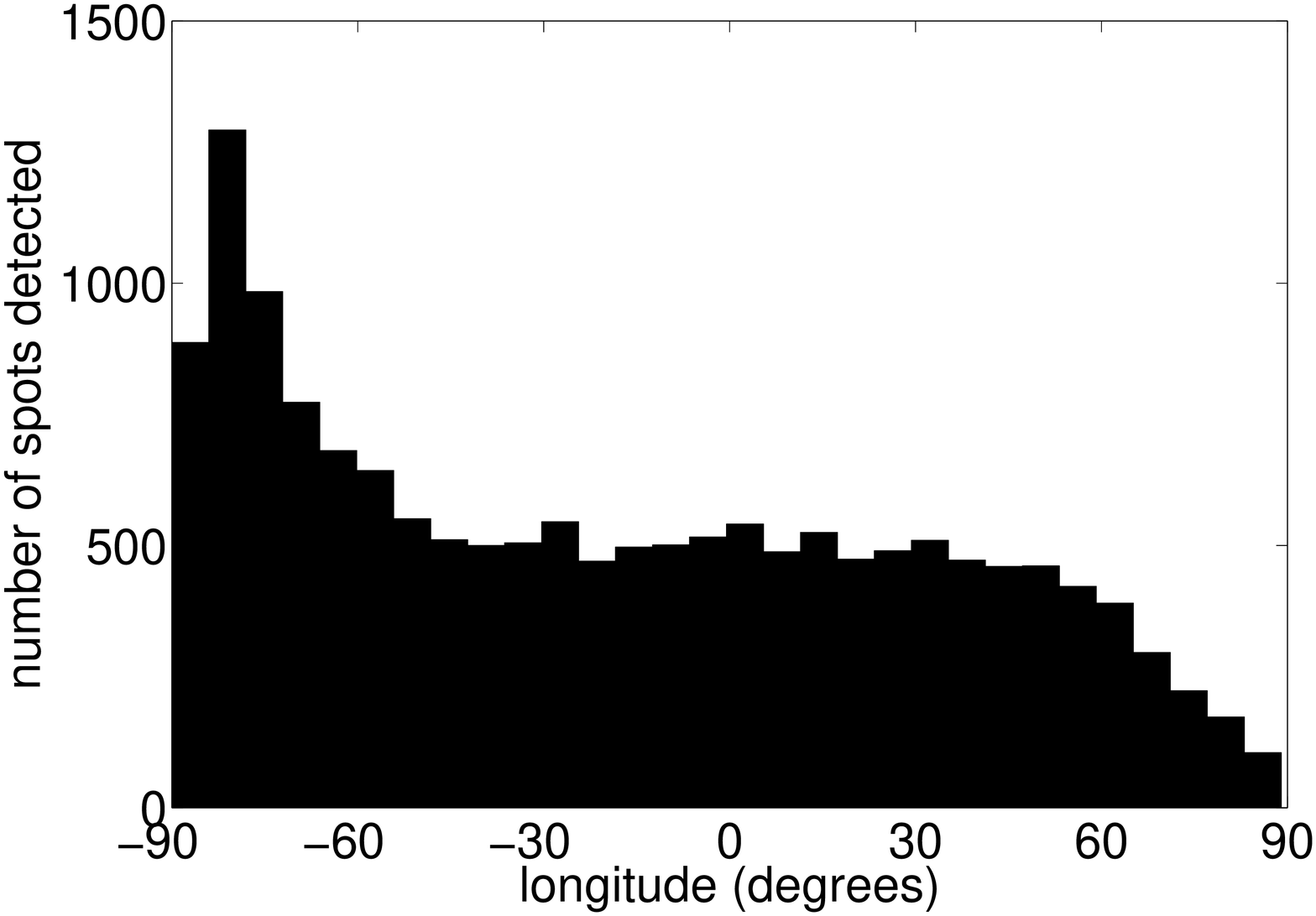} \\
(a) & (b) \\
\includegraphics[width=0.45\textwidth]{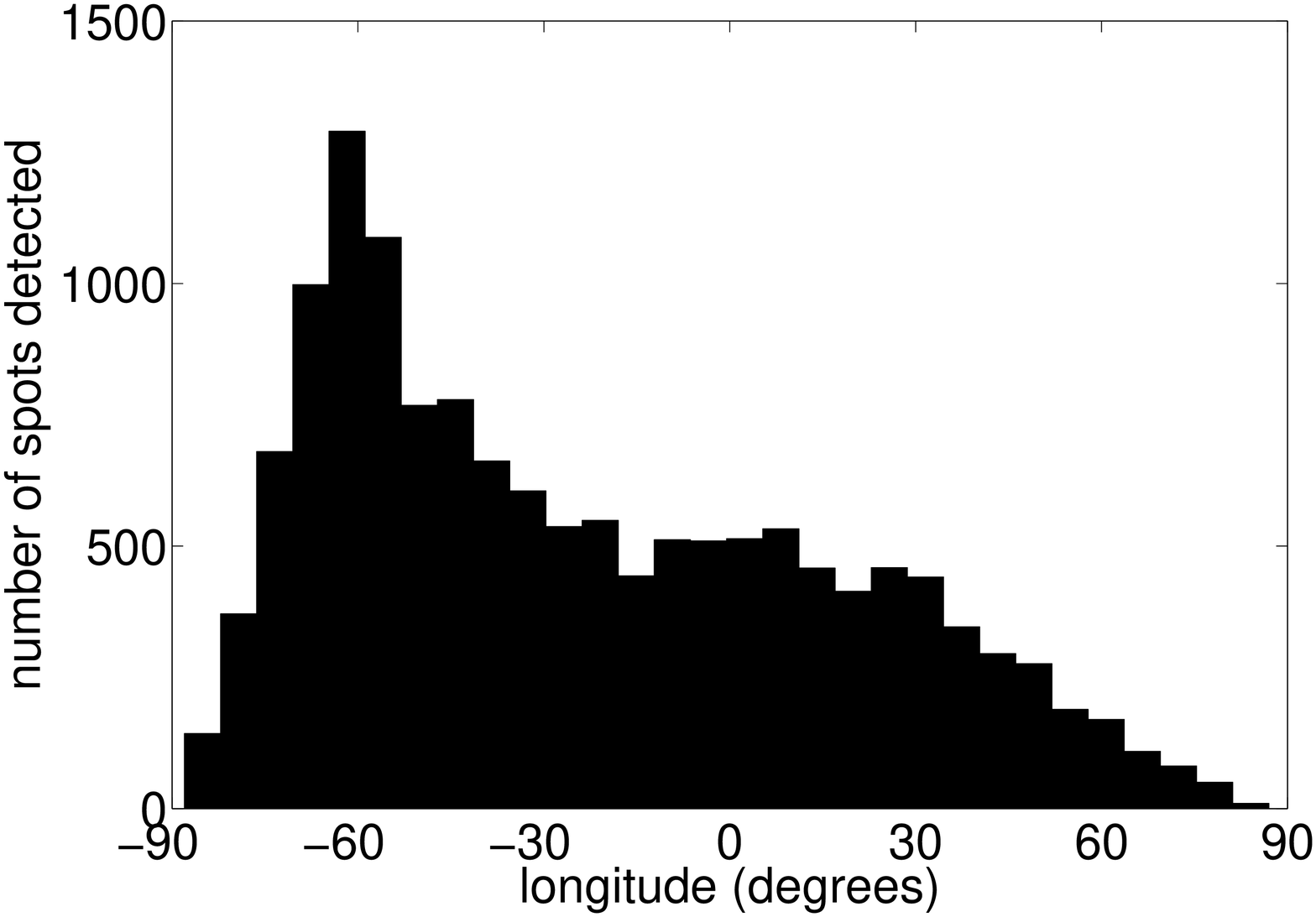} & \includegraphics[width=0.45\textwidth]{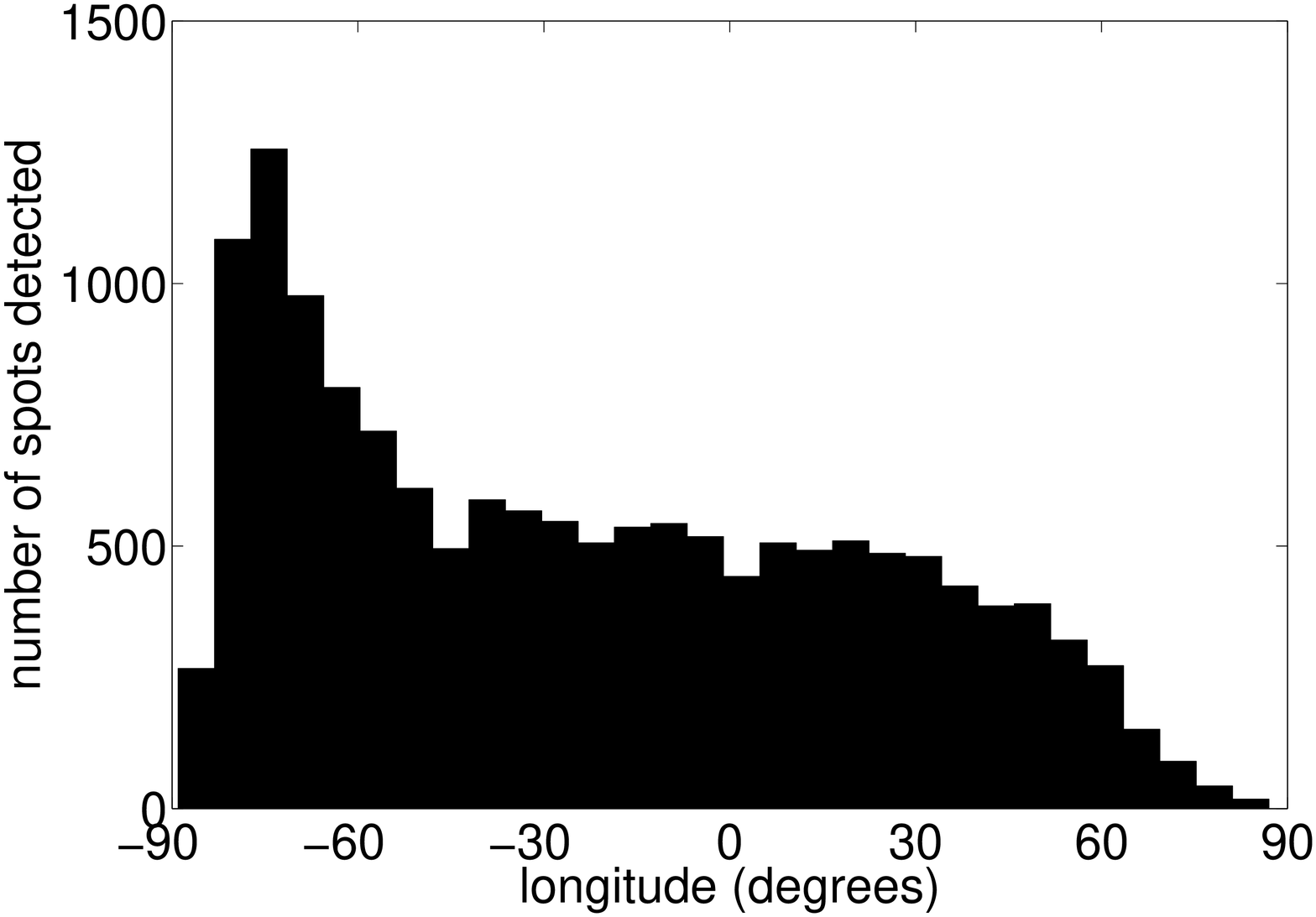} \\
(c) & (d)
 \end{tabular}
\caption{Distributions of sunspot emergences using different values for the power law of initial areas and growth rates. (a) power law with power -1.5 and growth rate of 17\% per 12 hours. (b) power law with power -1.5 and growth rate of 37\% per 12 hours. (c) power law with power -2.5 and growth rate of 17\% per 12 hours. (d) power law with power -2.5 and growth rate of 37\% per 12 hours.}\label{fig:parameters}
\end{figure}

The free parameters in the model are the initial distribution of spot areas and the rate at which the spots grow. To constrain these parameters, the initial spot area is always between 10MSH and 1000MSH in our model (MSH is millionths of the visible solar hemisphere) as it is rare to see a spot bigger than this (\opencite{2005A&A...443..1061B}) although sunspot groups this size are more common. The power law and growth rate are determined by testing different values and comparing the resulting distribution against the observations given in Figure~\ref{fig:histograms} using the KS test. The best fitting power law is an initial area distribution with $p = -2.5$ and the spots grow at a rate of 2\% of their area every 45 minutes for 5 days (these values were chosen as they correspond to intervals of the time between successive MDI continuum images and are equivalent to a growth rate of 37\% every 12 hours). If they have not yet reached the threshold to be visible at their solar longitude after 5 days, they can be removed from the sample. The effects of varying these parameters are shown in Figure~\ref{fig:parameters}.

The primary features of the observed distributions that we wish to replicate are the location of the main peak and the ratio between the height of the peak and the flatter region in the centre of the distribution. Panel (a) of Figure~\ref{fig:parameters}, representing the simulation with $p = -1.5$ and growth rate = 1\% every 45 minutes (equivalent to 17\% every 12 hours), shows the peak in the wrong place but in panel (b) we see that changing the growth rate from 17\% to 37\% every 12 hours moves the peak into better agreement with observations. However, the ratio between the height of the peak and height of the centre region is too low. By changing the power law from p = -1.5 to -2.5, in panel (c) we see the height of the peak increases relative to the centre region and provides a better comparison with the data obtained from SOHO (Figure~\ref{fig:histograms}, left panel). Combining these two parameters we see the best fitting model in panel (d) with $p = -2.5$ and growth rate = 37\% every 12 hours. This model gives a KS test $D$-value of 0.090.

\section{Monte-Carlo Testing against MDI Data, Including the Wilson Effect}~\label{sect:testing}
\begin{figure*}
\centering
\includegraphics[width=0.85\textwidth,clip=]{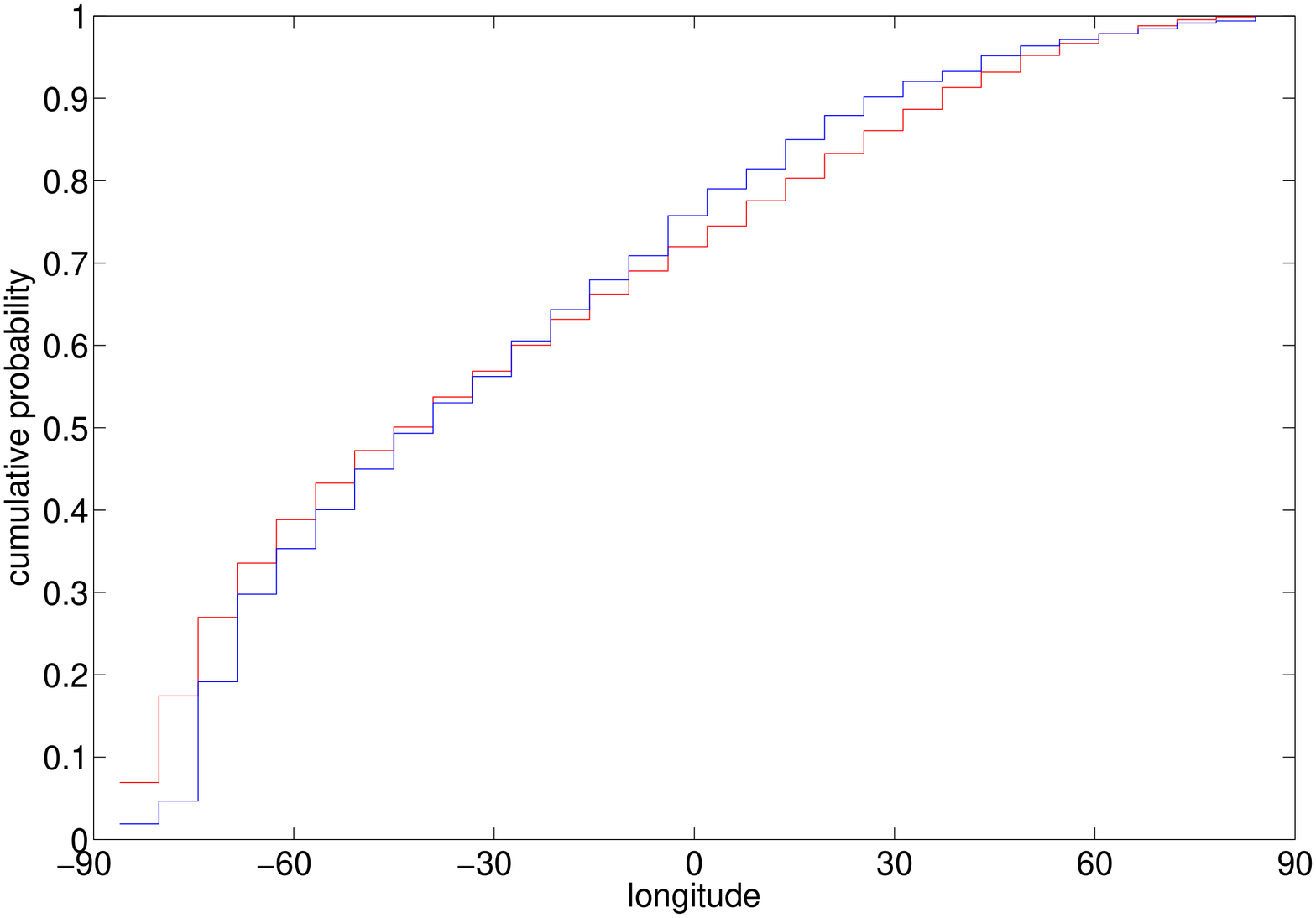}
\caption{The best fitting KS test without the effects of the Wilson depression included. A $D$-value of 0.090 was obtained for this test. The MDI observations are shown in blue, and the model in red. This fit was obtained with parameters $p = -2.5$ and growth rate = 37\% every 12 hours.}\label{fig:ksnowilson}
\end{figure*}

As described in Section~\ref{sect:monte}, searching the parameter space and using the KS test each time gave the parameters which best fit the MDI observations. These were initial spot areas that followed a power law distribution of power -2.5 and growth rate of 37\% in area every 12 hours for 5 days. Modelling these spot conditions, as well as the projection effects dependent on their position on the solar disk gave the KS test seen in Figure~\ref{fig:ksnowilson}. The comparison between the model and data is close enough to accept the null hypothesis that they were drawn randomly from the same larger set of data. However there is still clear room for improvement particularly in the high longitude regions where the peak of the sunspot emergence distribution is located. The importance of the Wilson effect increases as viewing angle increases and so this was added to the model in an attempt to obtain a better fit to observations, as described below.

\subsection{The Wilson Effect}~\label{sect:wilson}
\begin{figure*}
\centerline{\includegraphics[width=0.6\textwidth,clip=]{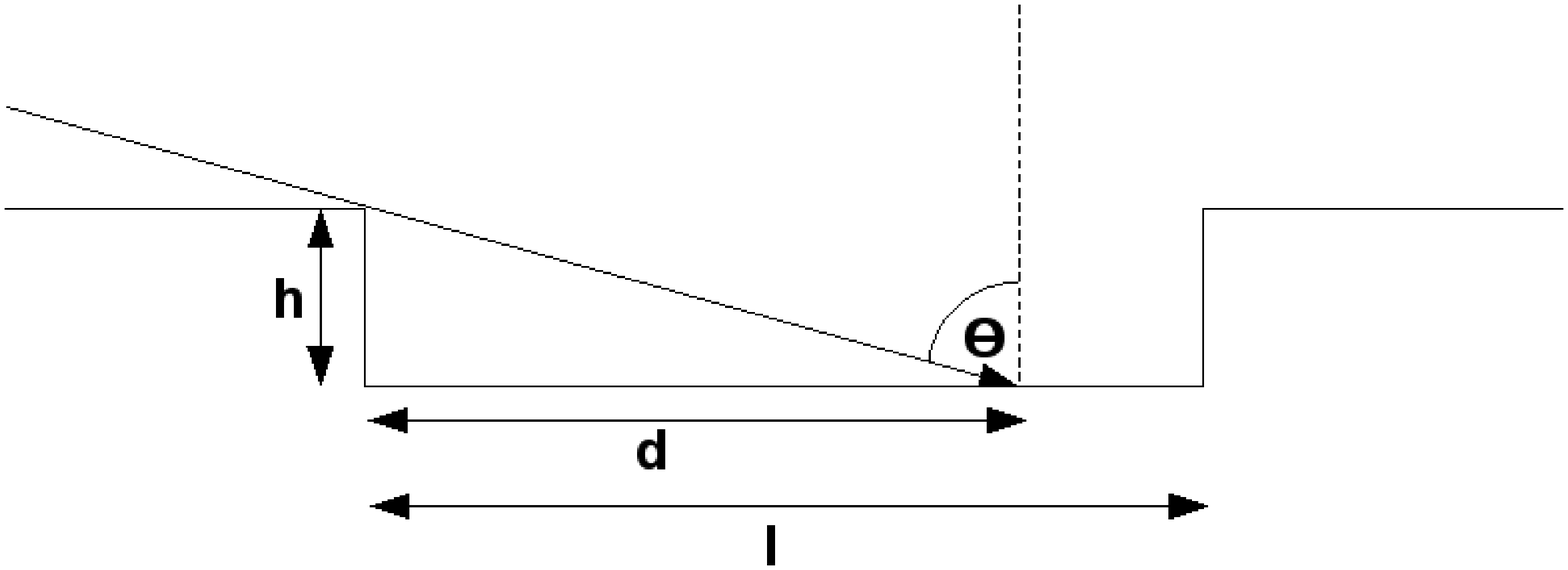}
            \hspace*{-0.03\textwidth}
	    \includegraphics[width=0.4\textwidth,clip=]{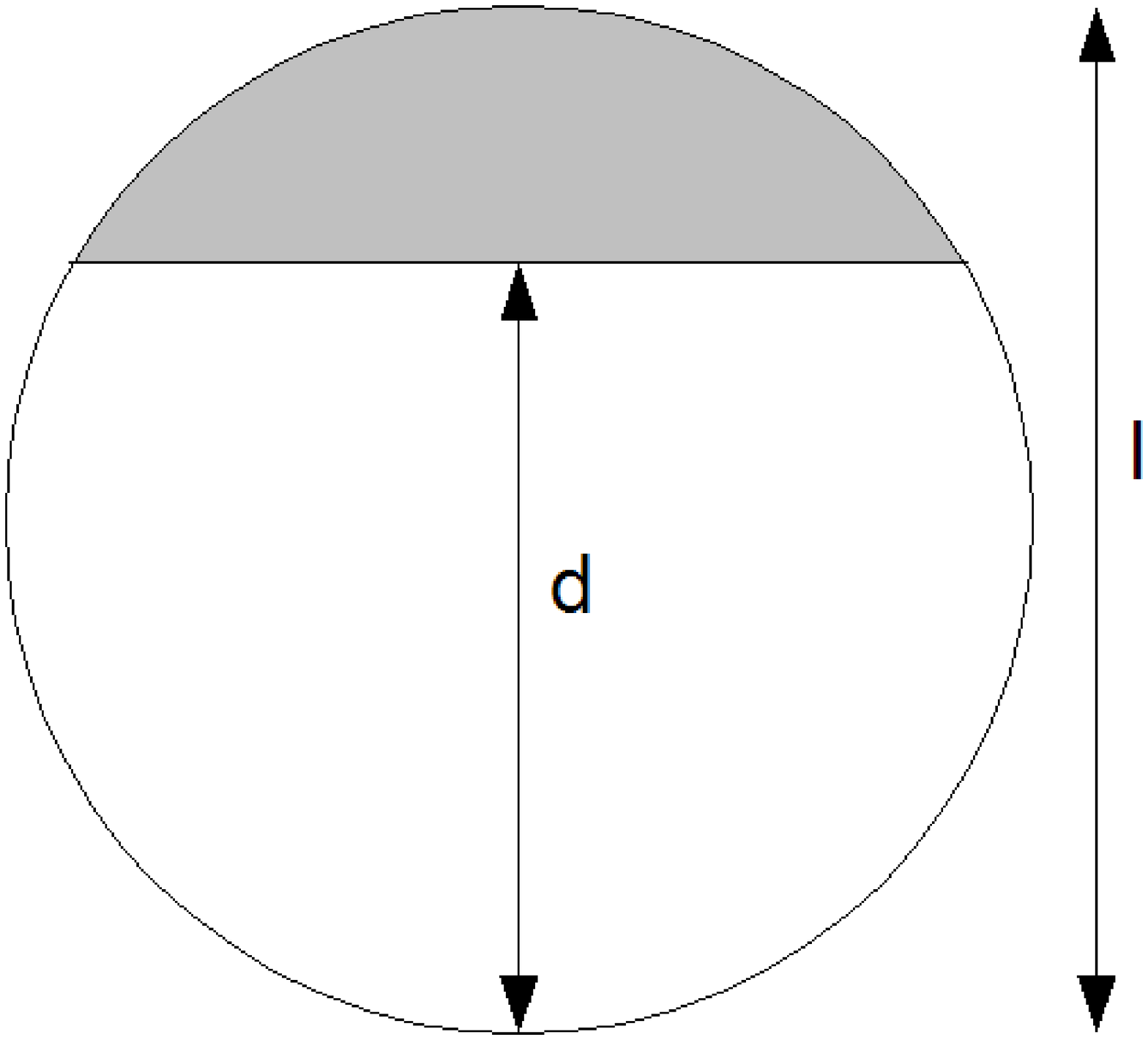}
	    }
\caption{The simplified model for a sunspot profile. On the left is a profile view of the sunspot and on the right is a view looking down into the sunspot along the normal to the solar surface.}\label{fig:wdcartoon}
\end{figure*}

The Wilson depression results in part of the sunspot being obscured by the surrounding photosphere. This can be thought of as a geometric effect, which will depend on the depth of the $\tau=1$ surface in the sunspot compared to the $\tau=1$ photospheric surface. To include the Wilson depression effect, a simplified geometric model of a spot was produced, and is shown in Figure~\ref{fig:wdcartoon}. The depression is modeled as a flat bottomed cylinder with vertical sides. When looking into a depression of this shape from a given angle to the normal to its base, $\theta$, there is a decrease in apparent diameter of the spot parallel to the line that passes through the spot between the centre of the solar disk and the limb. The apparent spot diameter is given by $l_{\mathrm{app}} = l - h \tan \theta$ where $h$ is the height of the walls that are obscuring the base, which has diameter $l$.

\begin{figure*} 
\centerline{\includegraphics[width=0.9\textwidth,clip=]{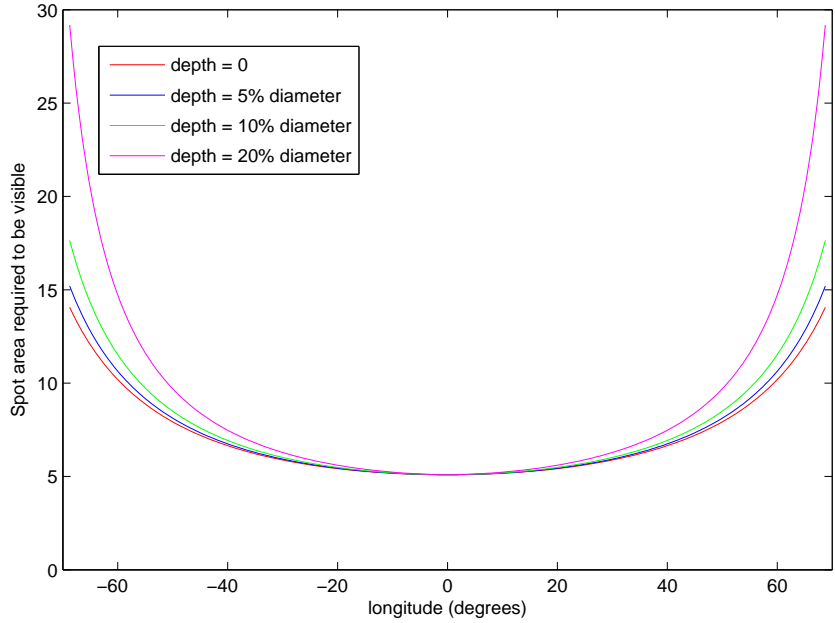}}
\caption{Visibility curves for models with no Wilson depression (depth = 0) and varying depressions. The plot only shows the range $\pm70^\circ$ as the visibility tends towards infinity at the limbs.}\label{fig:viscurves}
\end{figure*}

The visible area of the spot is then approximately the area shaded in the right panel of Figure~\ref{fig:wdcartoon} and can be given by the following equation for the area enclosed by a chord of a circle
\begin{equation}
 A_{\mathrm{chord}} = R^2 \cos^{-1} \left(\frac{R - l_{\mathrm{app}}}{R}\right) - (R - l_{\mathrm{app}}) \sqrt{2Rl_{\mathrm{app}} - l^2_{\mathrm{app}}}
\end{equation}
where $R = \frac{l}{2}$ is the radius of the disk. Hence, this area can be used in place of the original spot area, and all projection effects are applied to this area. Note here that the spot walls are not included in the visible area and we are only measuring the visible area of the spot base. As the spot diameter is much greater than the spot depth, this is a reasonable assumption. This modifies the visibility curve as shown in Figure~\ref{fig:viscurves} by making it more difficult to observe spots near the edges of the solar disk. This has the effect of increasing the visibility required at high longitudes. Only the range from $\pm70^\circ$ in solar longitude is shown as the $\cos(\theta)$ dependence of the visibility causes it to tend to infinity at the limbs. Even though the full $\pm90^\circ$ range is not shown, the effects of the Wilson depression are clear.

Again, a search of the parameter space can be performed but now including the Wilson depression effect in the model. The best fitting parameters are identical to that from the model without Wilson depression, in that an initial spot formation with areas obeying a power law with $p = -2.5$ and a growth rate of 37\% every 12 hours for 5 days returns the closest result to observations.

\begin{figure*} 
\centering
\includegraphics[width=0.85\textwidth,clip=]{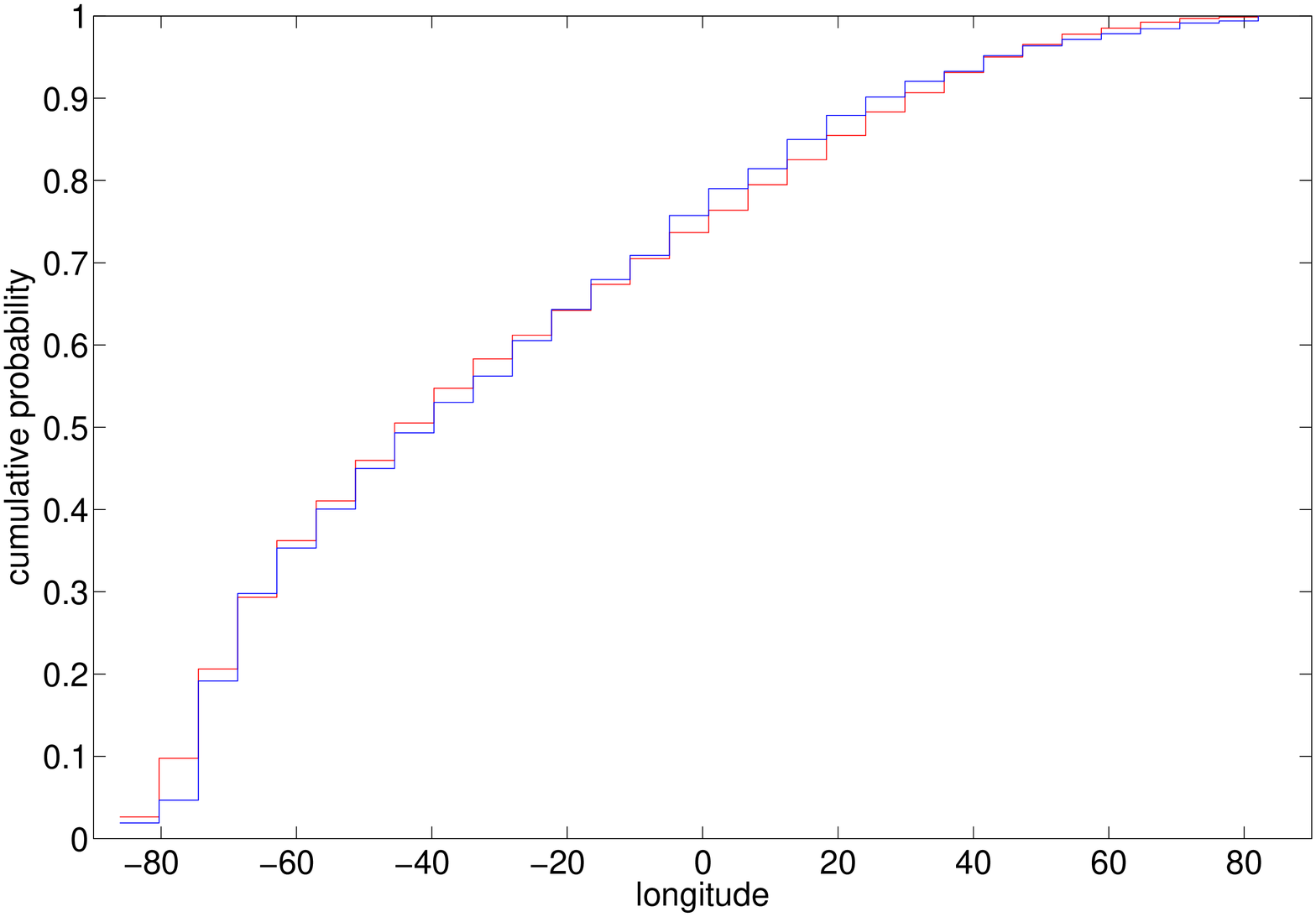}
\caption{The best fitting KS test with the effects of the Wilson depression included. A $D$-value of 0.061 was obtained for this test. The MDI observations are shown in blue, and the model in red.}\label{fig:kswilson}
\end{figure*}

The KS test for the best fitting model with the Wilson effect assuming a constant depth of 1000 km (\opencite{2003A&ARv..11..153S}) is shown in Figure~\ref{fig:kswilson}. This test returns a $D$-value of 0.061 which is a marked improvement over the best D-value without the depression effect included (note, smaller $D$-value is better). It is important to note that in this model, the depth of the sunspot umbra is the same regardless of spot size. A Wilson effect with depth proportional to spot size was also investigated but showed no consistent improvement over the model without the effect.

\subsection{Searching for an Optimum Spot Depth}~\label{sect:optimum}
\begin{figure}
\centering
 \begin{tabular}{c c}

\includegraphics[width=0.45\textwidth]{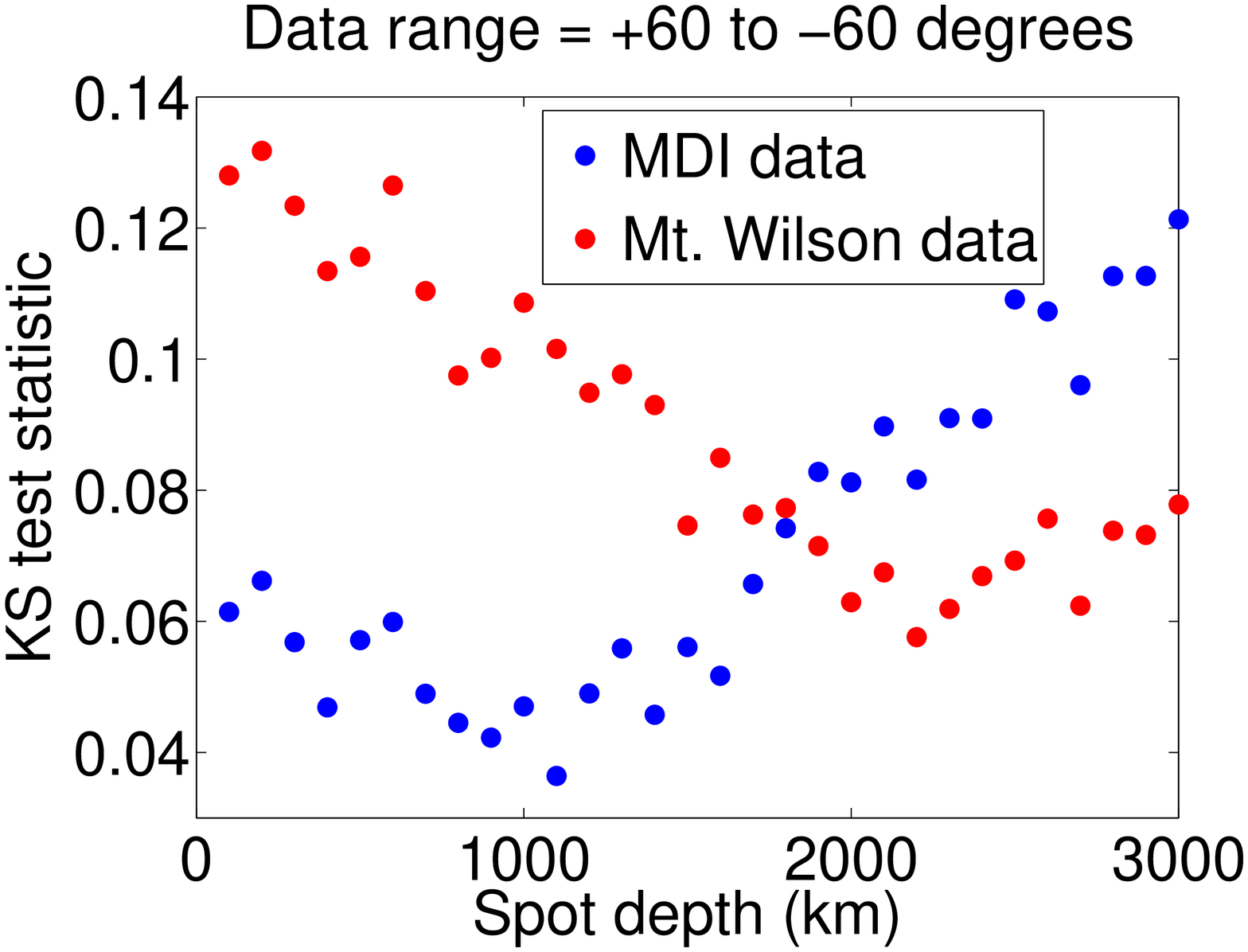} & \includegraphics[width=0.45\textwidth]{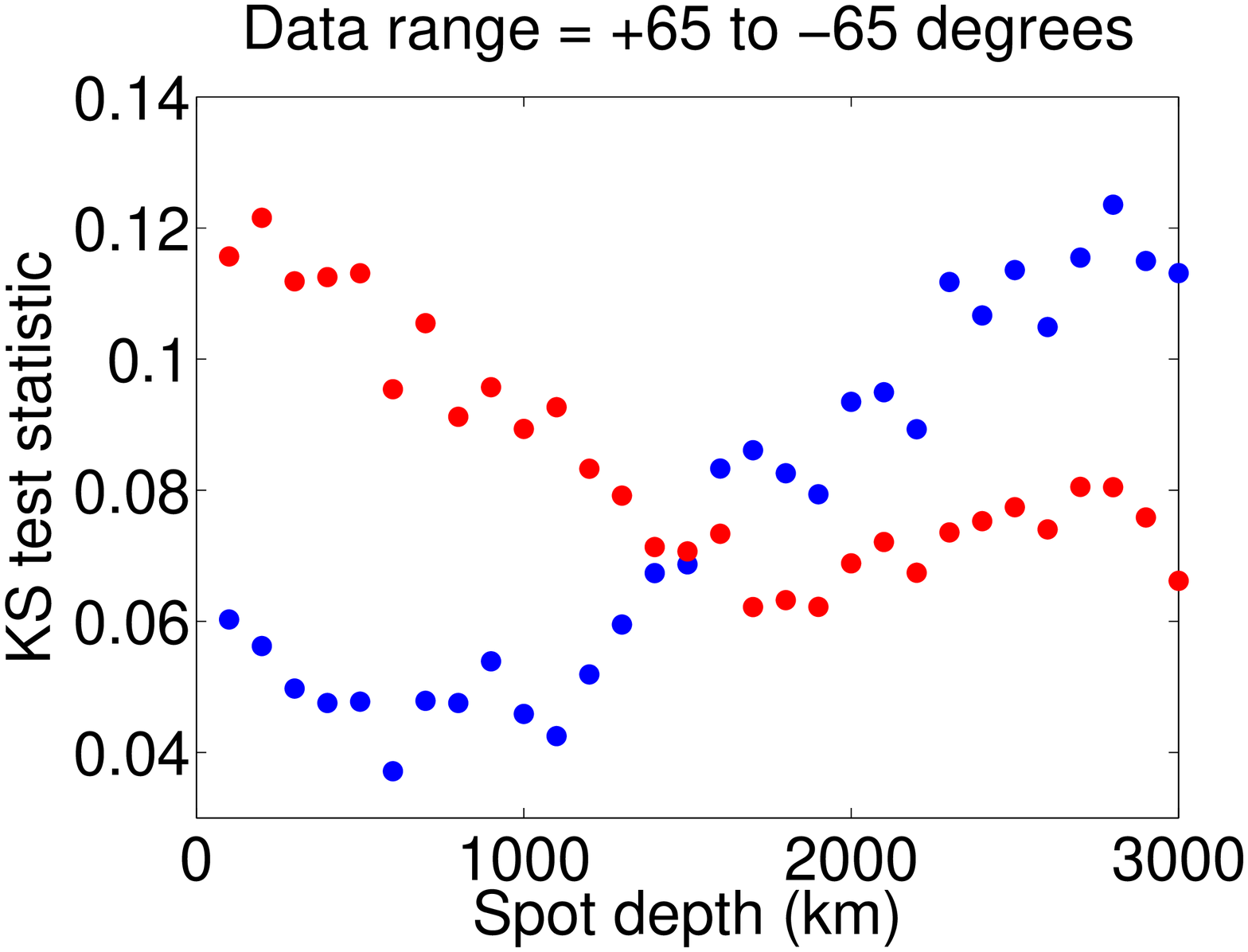} \\
\includegraphics[width=0.45\textwidth]{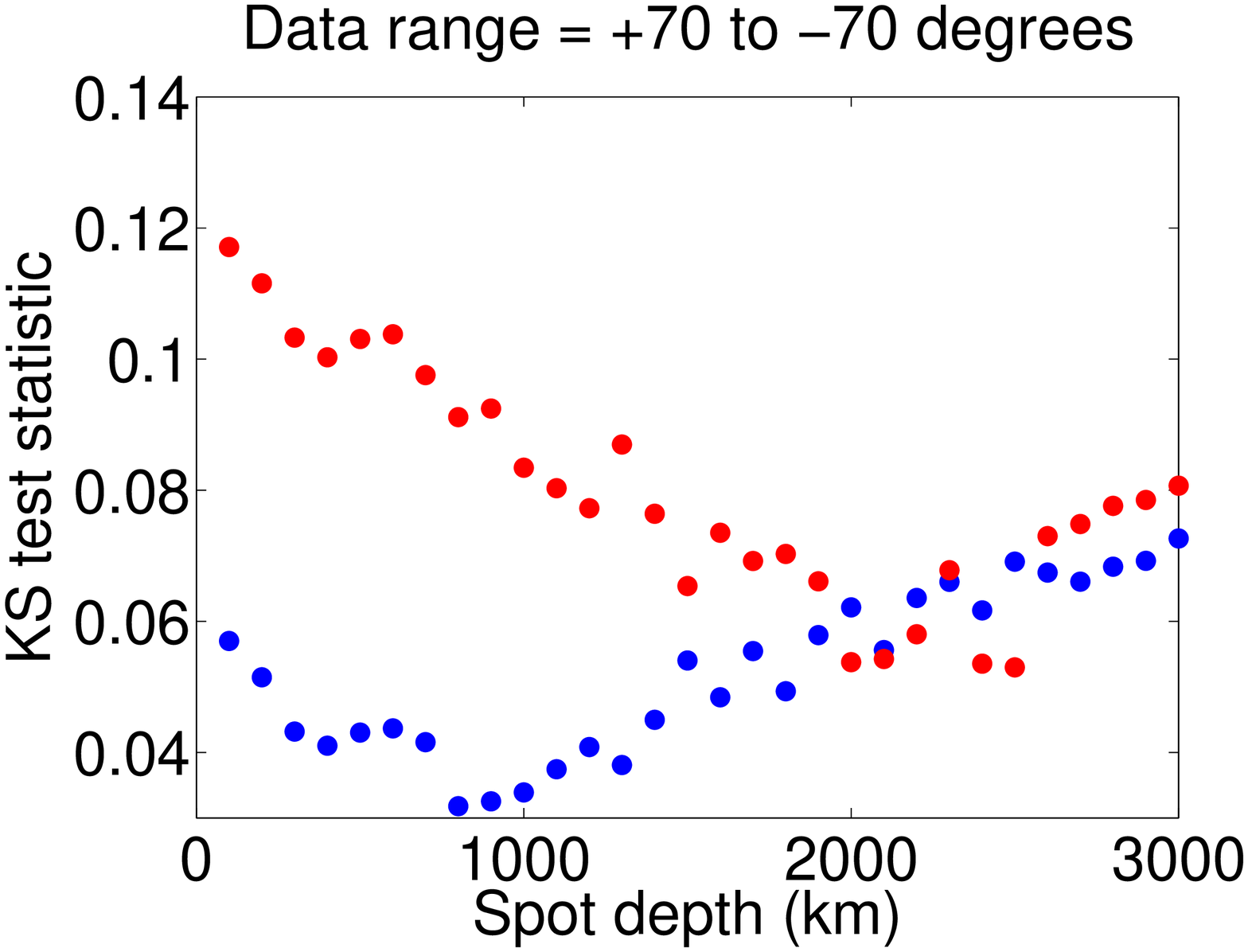} & \includegraphics[width=0.45\textwidth]{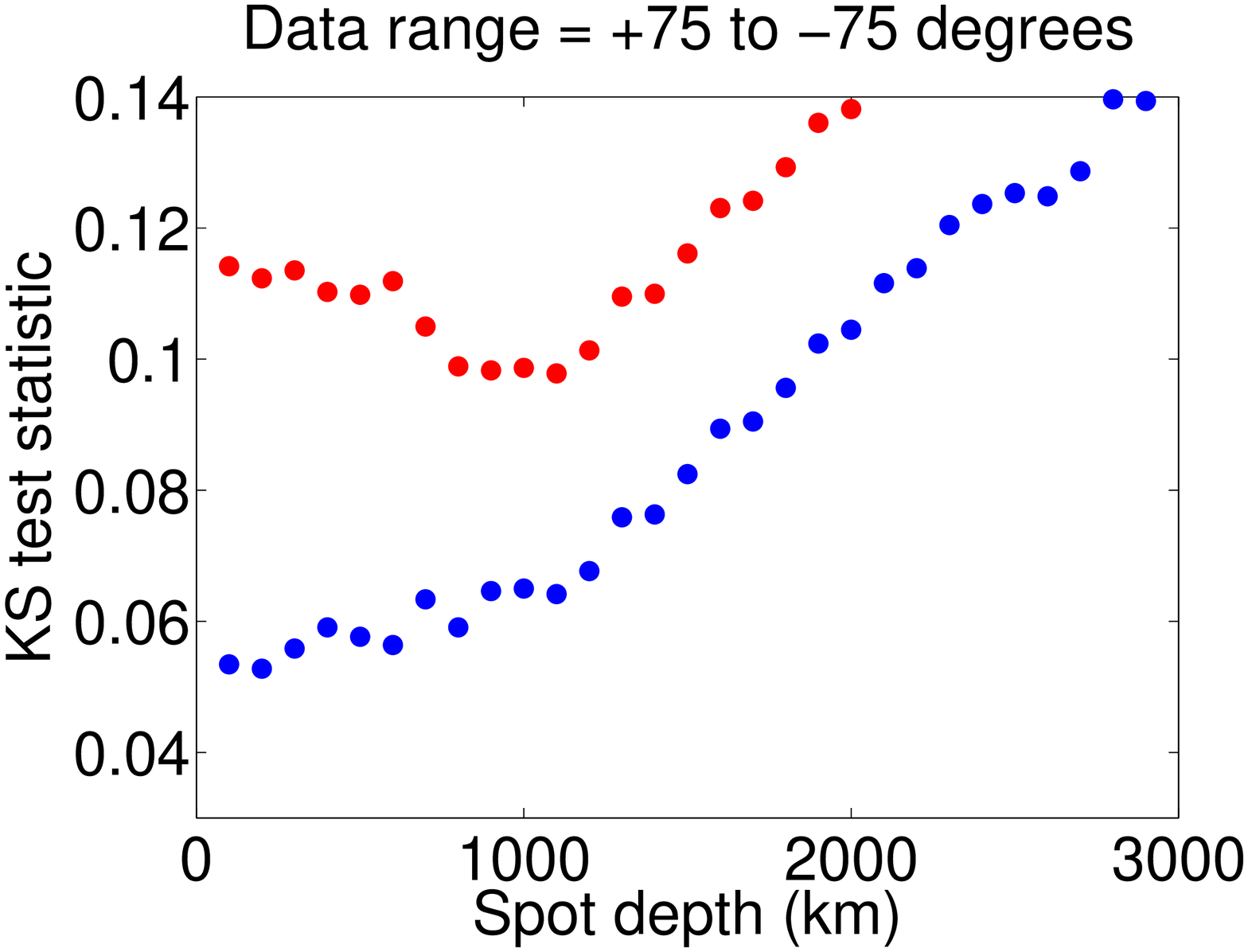} \\
\includegraphics[width=0.45\textwidth]{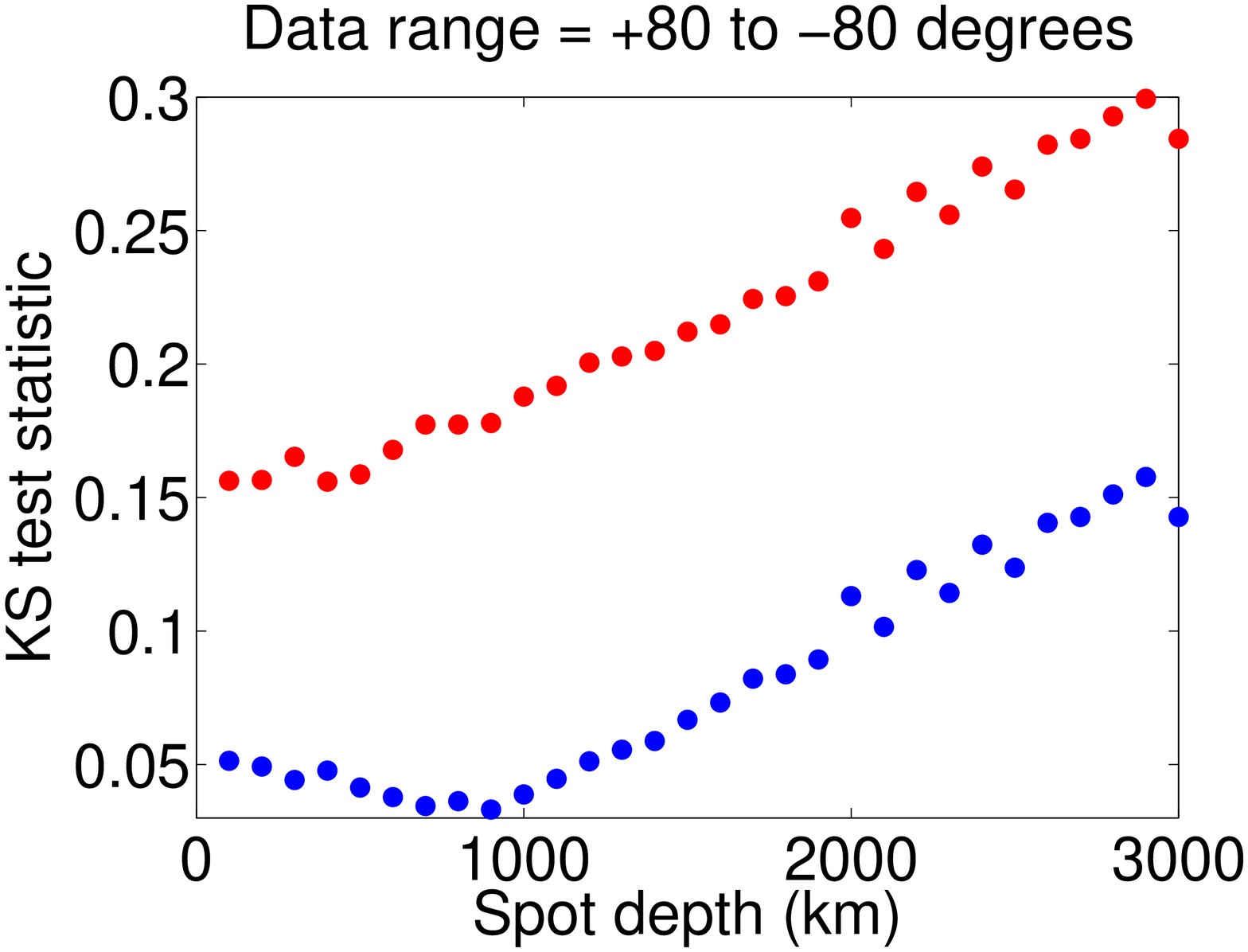} & \includegraphics[width=0.45\textwidth]{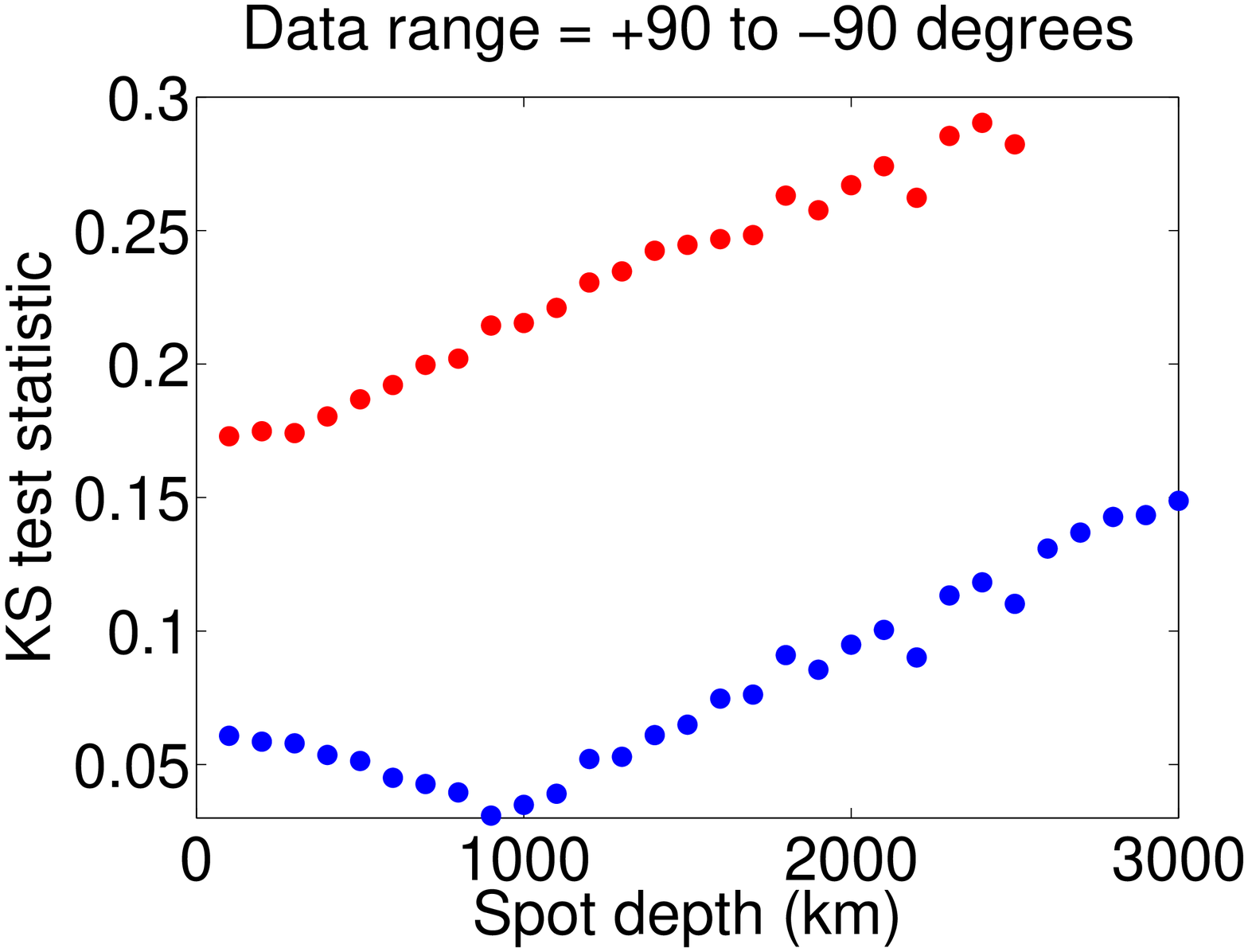}
 \end{tabular}
\caption{Series of KS tests for varying spot depths for a model including a power law distribution in initial spot areas. The MDI data appears to agree on a depth of 500-1500 km as the best fit to the model used, however the Mt. Wilson data does not agree on any particular value across the different data ranges.}\label{fig:KStests}
\end{figure}

If the Wilson depression model does indeed give an improvement over models that do not include the effect, there should be an optimum value for the spot depth that gives the best agreement with the observations. At this point, we also return to the Mt. Wilson data in our comparison. KS tests were performed comparing the MDI data and the Mount Wilson data with the model for spot depths varying from 0-3000km. Figure~\ref{fig:KStests} shows a plot of the KS $D$-value as a function of spot depth; the smaller the $D$-value the better the fit, so a minimum gives a range of optimum depths. Six different data ranges are plotted to determine whether the effects of including spots near the limbs changes the optimum depth of spots. The reason for investigating this is that the effect of the depression on the visibility curve is stronger as the spot approaches the limb (as seen in the visibility curves from Figure~\ref{fig:viscurves}), so there is more sensitivity to spot depth when data from near the limb are included. On the other hand, spots become less easy to detect in both manual and automated algorithms at high longitudes ($\pm 70^\circ$ and above) so there is presumably an optimum longitude range where we can take advantage of the shape of the visibility curve without being too troubled by difficulty in detection.

Note first that at zero spot depth (ie no Wilson depression) the $D$-value is always better for MDI than for Mt. Wilson. We interpret this as due to our algorithm detecting MDI spots and Mt. Wilson observers recording spot groups. This is described in more detail later.

Furthermore, the curves in Figure~\ref{fig:KStests} suggest that there is an optimum spot depth when comparing with MDI data. This spot depth is in the range of 500-1500 km and is in agreement with values given in \inlinecite{2003A&ARv..11..153S} of 400-2100 km. However, when we examine the Mt. Wilson data over the same ranges we do not see a consistent best value for the spot depth. A minimum in the curve can be seen when data near the limbs are neglected but when the limb data are included then the Wilson effect only worsens the fit compared to zero spot depth. To determine the cause for this discrepancy, we compare the methods of obtaining the raw data from SOHO and from Mt. Wilson.

\begin{figure}
\centering
 \begin{tabular}{c}

\includegraphics[width=0.7\textwidth]{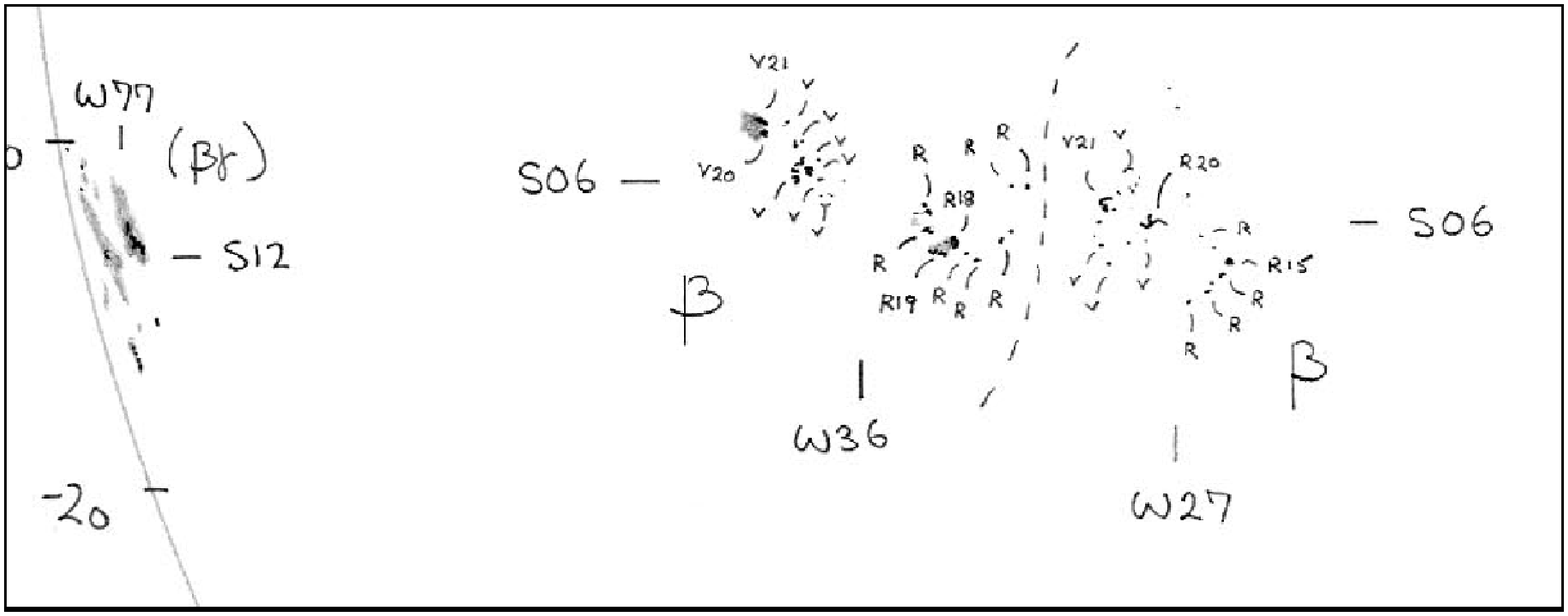} \\
\includegraphics[width=0.7\textwidth]{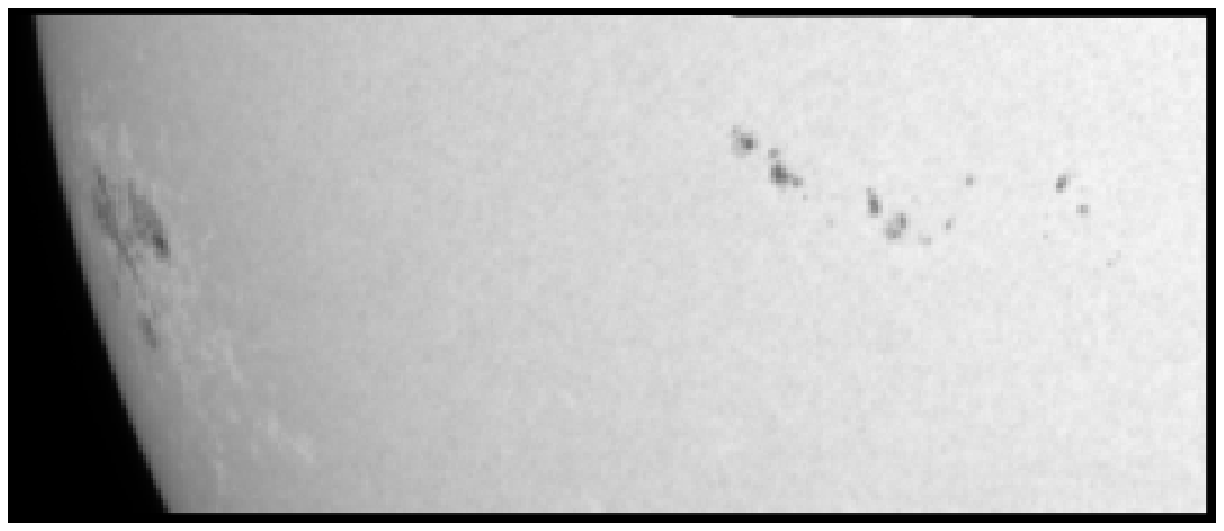} \\
\includegraphics[width=0.7\textwidth]{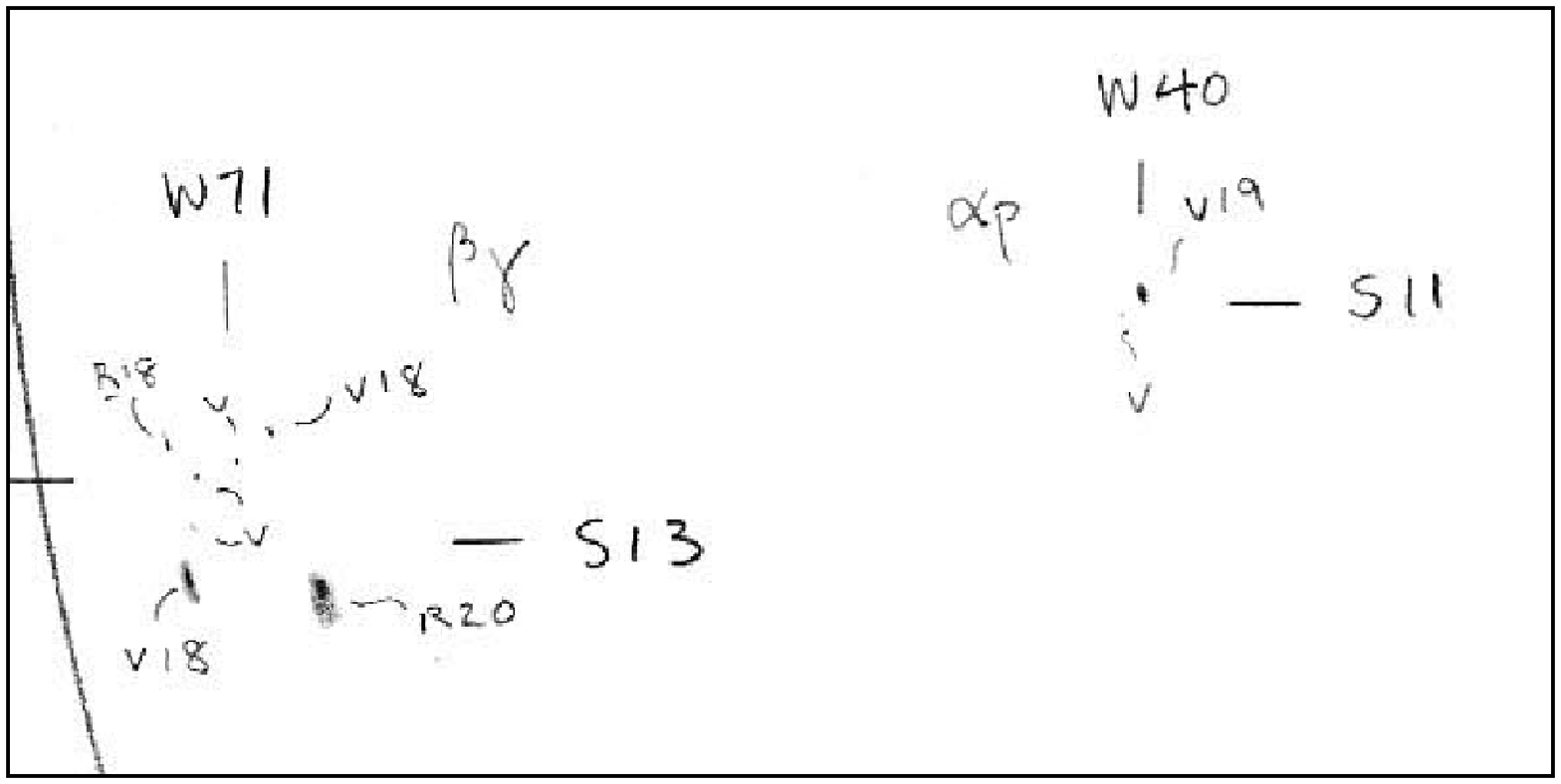} \\
\includegraphics[width=0.7\textwidth]{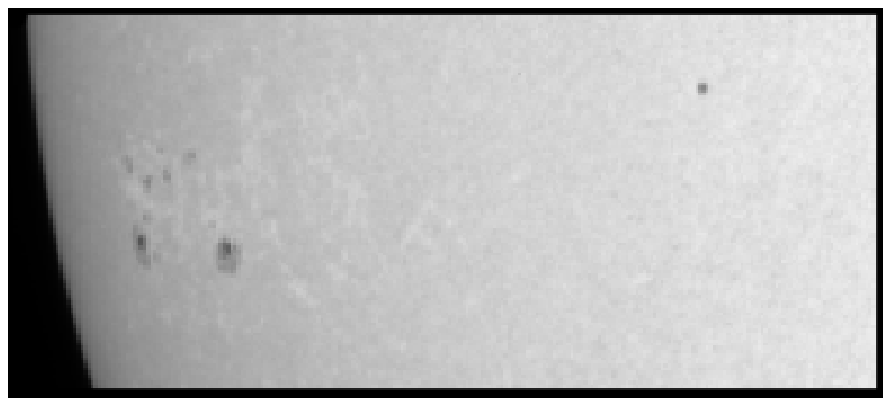}
 \end{tabular}
\caption{Sections from MDI continuum images and Mt. Wilson daily sunspot drawings showing an area near the solar limb with an active region present. The images are from (top pair) 26 September 2000 and (bottom pair) 6 April 2003. The co-ordinates on Mt. Wilson diagrams - \emph{e.g.} `W77, S12' are eye estimates of the area weighted position of group locations as determined by a human observer.}\label{fig:drawings}
\end{figure}

For this comparison we use sunspot drawings from Mt. Wilson from the dates of 26 September 2000 and 6 April 2003. These dates were chosen as there is an active region near the limb of the Sun in both cases and this allows us to demonstrate the effects of the Mt. Wilson `weighted' group positions in the most sensitive part of the visibility curve (ie around the peak in Figure~\ref{fig:histograms}, both panels). Blown up regions of the drawings are presented in Figure~\ref{fig:drawings} along with corresponding MDI images.

In the top panel, the sunspot group near the limb marked (S12, W77) is noted as a single group in the Mt. Wilson catalogue but our algorithm marks it as two separate sunspots with co-ordinates (S11, W70) and (S12, W75). The Mt.Wilson co-ordinates are determined by the observer and it should be noted that ``the location recorded is an eye estimate of the area-weighted position''. (The whole process can be found at \url{http://www.astro.ucla.edu/~obs/150_draw.html} for more detail). This effect is repeated in the lower panel where a single group is recorded in the Mt. Wilson catalogue at (S13, W71) and our algorithm returns a pair of spots at (S14, W65) and (S13, W73). As the data from Mt. Wilson is measuring active regions as opposed to individual sunspots we should not expect to see the same trends when we are analysing the Wilson effect because the datasets are recording different numbers of candidates in different positions. Note also how rapidly the visibility curves vary in Figure~\ref{fig:viscurves} at $|\lambda| > 70^\circ$ and how much they differ from one another for different Wilson depression depths. The $|\lambda| > 70^\circ$ region is therefore where most of the discriminatory power of this method is concentrated and the difference in candidates between the two sets of data in this critical region are what give the substantial difference between MDI and Mt. Wilson data presented in Figure~\ref{fig:KStests}. In addition, modelling the Wilson effect applied to an active region does not have as much physical meaning as applying it to a single sunspot. Although many active regions are indeed a single spot, active regions such as those seen in the upper panel of Figure~\ref{fig:drawings} (at co-ordinates (S06, W27) and (S06, W36)) contain many small spots. When analysing this region with our algorithm, we find four sunspots that are large enough to pass the cutoff filter of 30 MDI pixels and the spots are all treated with the Wilson effect separately.

For these reasons, the MDI data is telling us more about the properties of individual sunspots than the Mt. Wilson data. It should be noted that the sunspot emergence plots shown at the start of the paper in Figure~\ref{fig:histograms} are similar, since sunspots and active regions are both detected in the same physical locations but the effectiveness of the Mt. Wilson data breaks down when attempting to use the same data to ascertain the depth of sunspots due to the Wilson effect. This is because the visibility curve is very sensitive at high longitude values, whereas the `averaged' appearance positions for entire groups at these longitudes are a relatively poor approximation for actual spot emergence positions.

For completeness, the analysis of the model is repeated here using a lognormal distribution in spot areas rather than the power law treatment presented earlier. \inlinecite{1988ApJ...327..451B} and \inlinecite{2005A&A...443..1061B} find that a lognormal distribution in sunspot areas agrees with the observations, however they are not looking specifically at the area distribution \emph{at emergence}.

\begin{figure}
\centering
 \begin{tabular}{c c}

\includegraphics[width=0.46\textwidth]{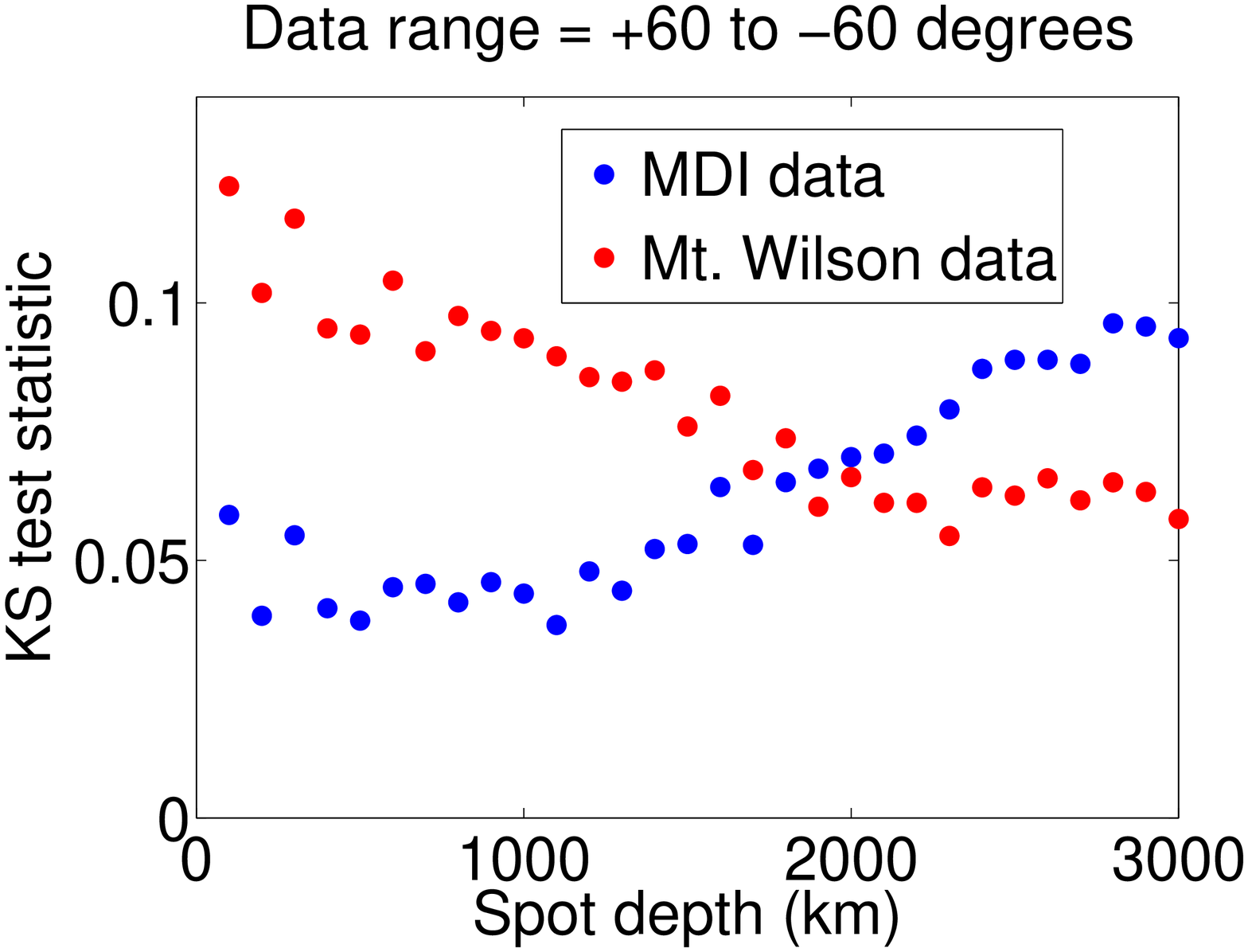} & \includegraphics[width=0.46\textwidth]{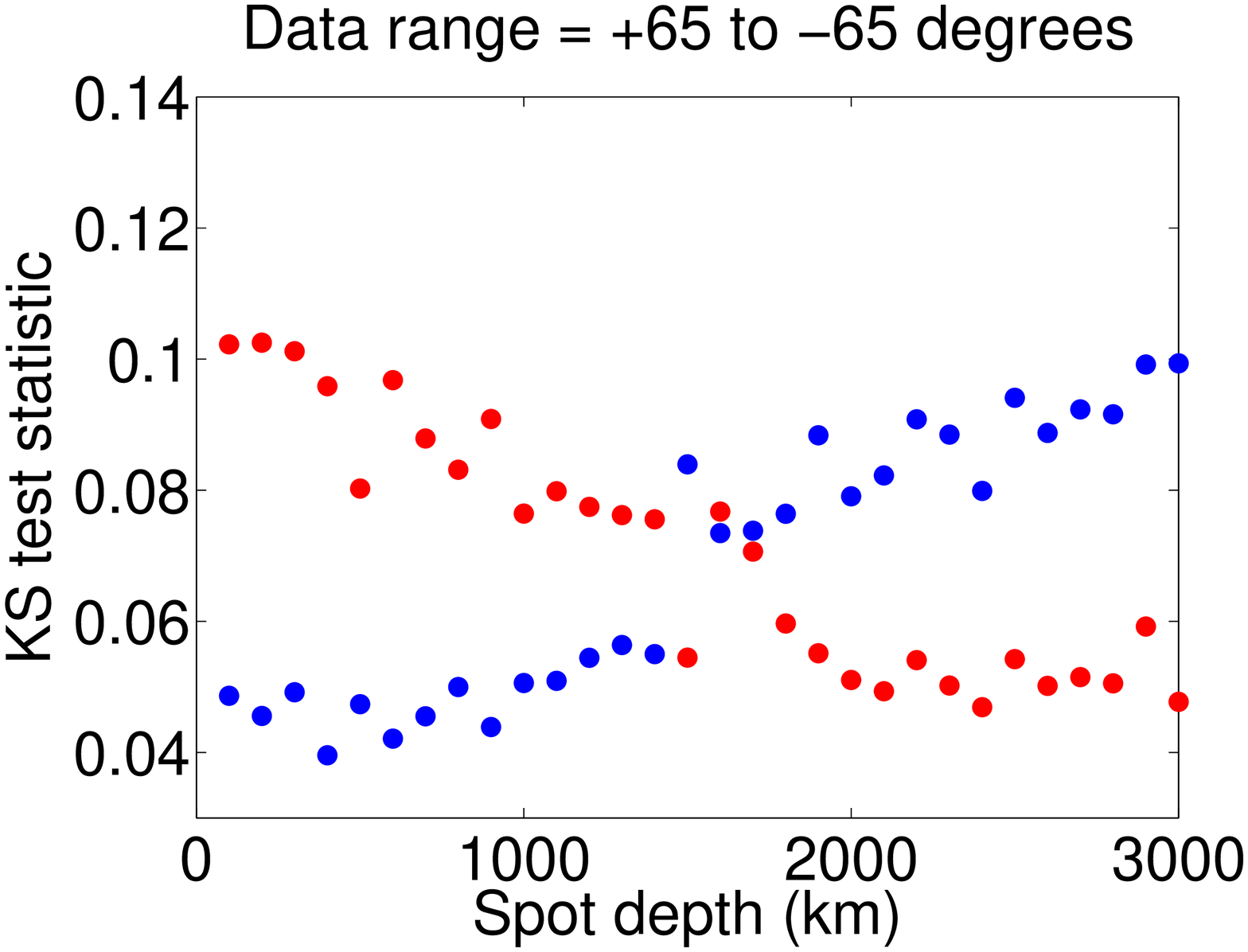} \\

 \end{tabular}
\caption{Series of KS tests for varying spot depths for a model including a lognormal distribution in initial spot areas. The left panel includes spots first detected in longitude range $\pm60^\circ$ and the right panel includes spots first detected in longitude range $\pm65^\circ$}\label{fig:logKStests}
\end{figure}

We present the KS test series shown in Figure~\ref{fig:logKStests} for the lognormal distribution in spot area at emergence. Only the tests for $\pm60^\circ$ and $\pm65^\circ$ in spot longitude are given as these have the best quality of data due to the visibility effects as spot longitude increases. These can be compared to the KS test series with the power law distribution in spot area in Figure~\ref{fig:KStests}.

These plots show similar features to those shown in Figure~\ref{fig:KStests} but they are not identical. We see when a lognormal distribution is taken in the model, we do not get as big a difference between the $D$-value at depth = 0 and the minimum $D$-value. Also, when looking at the $\pm65^\circ$ data (Figure~\ref{fig:logKStests}, right panel) it is not clear whether there is a minimum of the curve that is at a spot depth $\neq$ 0. The differences seen in the result when changing the areas at emergence show us the importance of determining the true sunspot area distribution at emergence from observation and we leave this for future work.

\section{Discussion and Conclusions}~\label{sect:disc}
The algorithm created to detect sunspots from the MDI continuum data was both quick and accurate when building a catalogue of sunspot emergences. This is supported by the agreement with the distribution found by \inlinecite{2008A&A...479L...1D}. A sunspot emergence model was then constructed to determine whether the observed emergence distribution could be explained by foreshortening effects alone. A model with parameters of a growth rate of 37\% every 12 hours and an initial spot area distribution with power index $p = -2.5$ was found to best fit the data although there was room for improvement. The Wilson depression effect was included in the model using the geometric treatment described in this paper and was found to give a closer fit to the data. To determine the fit, the KS test was used.

As including the Wilson depression effect produced an improvement in the model fit, a search for the value of spot depth which best fitted the data was undertaken. This provided two conflicting sets of results depending on which data set was used. For the MDI data an optimum value for the sunspot depth of 500-1500 km was obtained (which is in agreement with the values given in \opencite{2003A&ARv..11..153S}) whereas the Mt. Wilson data showed no consistent optimum value. This was repeated using a lognormal distribution in sunspot areas at emergence and similar trends were found to those using the power law distribution although observations of the true area distribution at emergence would be valuable in constraining the choice of model. After examining how the data was captured it was determined that because Mt. Wilson recorded average active region positions instead of sunspots, applying an average Wilson depression depth to an `averaged' spot position (representing some mean property of the active region) is a poor approximation in a region where the visibility curve is changing so fast. This was compounded by the fact that our algorithm had greater sampling power near the limbs of the Sun as we could examine each individual spot separately.

We also considered the effects of a model in which the spot walls are not vertical as shown in this paper, but sloped and reach the spot base at a distance $l/4$ from the spot centre. We concluded that this would not have a significant effect on the results we find. This is due to the viewing angles at which the spot base can be obscured by the spot walls. For the model given in this paper, obscuration happens at all angles (other than at disk centre) and this obscuration effect is greater at angles above about $60^\circ$ (see Figure~\ref{fig:viscurves}). For a model with sloped walls, there is no obscuration until the viewing angle is greater than the slope of the walls (which is $\mathrm{tan}^{-1}(l/4h)$ for this geometry). This limits the effectiveness of the diagnostic to angles greater than $75^\circ$ - $80^\circ$ for spots large enough to be detected and assuming a Wilson Depression value in line with that found by other means and in the literature (Solanki (2003) for example). At angles less than this critical angle, the `near side' penumbra will still be visible and so the spot umbra will not be obscured.

At angles greater than this critical angle, the obscuration of the spot base by the `near side' penumbral wall is similar to the model used in this paper. However, there is another critical angle in the sloped wall geometry and this occurs when the whole spot base (or umbra) is obscured and the `far side' penumbra is still visible. It is above this angle that the models should show the greatest difference. For this geometry, this second critical angle is $84^\circ$ - $85^\circ$ and at these solar longitudes, geometric foreshortening causes only the largest of spots to be detectable. As these largest of spots form less than 3\% of the total distribution this will limit the effect on the depth that we deduce.

This leads us to conclude that the simple spot model adopted here is adequate for determining the depth of the Wilson depression of sunspots.

We also conclude that the Wilson depression effect does improve our model when compared against SOHO MDI continuum data and also that a spot depth of 500-1500 km gives the best fit to the observations for the simple model used.

%

%
 \begin{acks}
FW acknowledges the support of an STFC PhD studentship. LF acknowledges support of STFC Rolling Grant no. ST/F002637 and EC support via the SOLAIRE RTN (MTRN-CT-2006-035484). This study includes data from the synoptic program at the 150-Foot Solar Tower of the Mt. Wilson Observatory. The Mt. Wilson 150-Foot Solar Tower is operated by UCLA, with funding from NASA, ONR and NSF, under agreement with the Mt. Wilson Institute. SOHO is a project of international cooperation between ESA and NASA. We would like also to thank our anonymous referee for constructive comments which have improved the paper.
 \end{acks}

%
%
 \bibliographystyle{spr-mp-sola}
 \bibliography{asymmetry_V2}  
%
%
%
%

\end{article} 
\end{document}